\documentstyle[preprint,prl,aps]{revtex}

\tighten

\def\beq{\begin{equation}}
\def\eeq{\end{equation}}
\def\beqa{\begin{eqnarray}}
\def\eeqa{\end{eqnarray}}
\def\lla{\left\langle}
\def\rra{\right\rangle}
\def\za{\alpha}
\def\zb{\beta}
\def\lsim{\mathrel{\raise.3ex\hbox{$<$\kern-.75em\lower1ex\hbox{$\sim$}}} }
\def\gsim{\mathrel{\raise.3ex\hbox{$>$\kern-.75em\lower1ex\hbox{$\sim$}}} }

\begin{document}
\draft
\preprint{{\vbox{\hbox{IPAS-HEP-k008}\hbox{NCU-HEP-k004} \hbox{Jun 2002} 
\hbox{ed. Mar 2004}}}}


\title {On the Formulation of the Generic Supersymmetric Standard Model
(or Supersymmetry without R parity) }
\author{\bf Otto C. W. Kong}
\address{Department of Physics, National Central University, Chung-li, TAIWAN 32054
\\
Institute of Physics, Academia Sinica, Nankang, Taipei, TAIWAN 11529
}
\maketitle

{\tighten
\begin{abstract}
The generic supersymmetric version of the Standard Model would 
have the minimal list of superfields incorporating the Standard Model
particles, and a Lagrangian dictated by the Standard Model gauge 
symmetries. To be phenomenologically viable, soft supersymmetry breaking 
terms have to be included. In the most popular version of the supersymmetric 
Standard Model, an {\it ad hoc} discrete symmetry, called R parity, is added in 
by hand. While there has been a lot of various kinds of R-parity violation
studies in the literature, the complete version of supersymmetry
without R parity is not popularly appreciated. In this article, we
present a pedagogical review of the formulation of 
this generic supersymmetric Standard Model and give a detailed 
discussion on the basic conceptual issues involved. Unfortunately,
there are quite some confusing, or even plainly wrong, statements on
the issues within the literature of R-parity violations.  We aim at 
clarifying these issues here.  We will first discuss our formulation, about 
which readers are urged to read without bias from previous acquired 
perspectives on the topic. Based on the formulation, we will then 
address the various issues . In relation to phenomenology, our review here will 
not go beyond tree-level mass matrices. But we will give a careful discussion
of mass matrices of all the matter fields involved. Useful expressions for 
perturbative diagonalizations of the mass matrices at the phenomenologically 
interesting limit of corresponds to small neutrino masses are derived. All these
expressions are given in the fully generic setting, with information
on complex phases of parameters retained. Such expressions have
been shown to be useful in the analyses of various phenomenological
features. 
\end{abstract}
}


\section{Introduction}
The minimal supersymmetric standard model (MSSM) is no doubt the most
popular candidate theory for physics beyond the Standard Model (SM). An
alternative version with a discrete symmetry, called R parity, not imposed
deserves no less attention. {\it Supersymmetry without R parity is nothing 
but the generic supersymmetric Standard Model\cite{as12}, {\it i.e.}, a 
theory built with the minimal superfield spectrum incorporating the SM 
particles, interactions dictated by the SM (gauge) symmetries, and the idea 
that supersymmetry (SUSY) is softly broken (at or below the TeV scale).}
$\!\!$\footnote{This review is meant to be pedagogical. Only a minimal basic 
background of supersymmetry is indeed presumed. The rest of the 
discussions in this section set our case among the background literature 
of supersymmetric Standard Model(s). Readers not familiar with the latter
or the notion of R parity may want to skip that. The statement in italics
above {\it defines} the model we discuss here.  In fact, some idea about what 
the statement means is all that is needed to appreciate the present review. 
}
 
From the theoretical point of view, R parity is simply 
an {\it ad hoc} global symmetry imposed by hand. It does simplify the
Lagrangian very substantially and restores the accidental symmetries of
baryon and lepton numbers of the SM, but is not otherwise well motivated.
Phenomenologically, not imposing R parity would beg an alternative 
mechanism to protect proton decay; and may resulted in, but not necessary 
mandates, losing the so-called lightest supersymmetic particle (LSP) as the 
favorite dark matter candidate. Concerning protecting proton decay, it has 
been established that R parity is not the only candidate for the job; nor is it
the most effective\cite{pd}. It is the most restrictive, though, in terms
of what terms are admitted in the renormalizable Lagrangian or otherwise. 
On the other hand, giving up R parity {\it does} allow the neutrinos to have 
masses and mixings {\it without} the need of introducing extra superfields 
beyond the supersymmetric SM spectrum. At the present time, experimental 
results from neutrino physics\cite{nu} is actually the only data we have 
demanding physics beyond the SM, while signals from SUSY are still 
absent\cite{g-2}. Hence, the case for giving up R parity is stronger than ever. 
The generic supersymmetric Standard Model (GSSM) is, at least conceptually, 
the simplest model with SUSY and neutrino masses. It also promises 
exciting new phenomenology, in collider machines and beyond, and a 
strong link between neutrino physics and the latter. Hence, we conclude
that it makes sense to take the GSSM and study the experimental constraints 
on the various couplings without {\it a priori} bias. From the theoretical 
point of view, in relation to proton decay, baryon number is expected to be 
protected by some sort of symmetry, while lepton numbers have to be
violated.

There are certainly no lack of studies on various ``R-parity violating models" 
in the literature. However, such models typically involve strong assumptions
on the form of R-parity violation. In most cases, no clear statement on
what motivates the assumptions taken is explicitly given. In fact, there are 
quite some confusing, or even plainly wrong, statements on issues concerned.
It is important to distinguish among the different R-parity violating (RPV) 
``theories", and, especially, between such a theory and the unique GSSM. 
Results from studies on a  particular version of RPV model would have no
general validity, if they are based on naive but strict assumptions on the
vanishing of a set of the generally admissible RPV couplings.
However, by carefully embedding such studies into the GSSM framework, 
and addressing explicitly the questions of how small the neglected couplings
have to be in order not to upset the phenomenological features under
study, we could piece together a more comprehensive story of interesting
GSSM phenomenology.  

Many works on RPV phenomenology give the wrong claim or lead to the wrong
impression that they are studying some aspects of the GSSM. The truth is that 
the strong assumptions, hidden or otherwise, behind their formulations 
restrict the general validity of their studies. There are subtle issues involved
in taking a specific RPV model as a limiting case of the GSSM. Such issues 
are important in the latter interpretation, but too often overlooked.  
We hope to clarify such issues in this review. 

If one looks at the GSSM independent of the
MSSM and the other RPV theories, the notion of R parity is simply not there at
all. In fact, to the extent that the physical leptons contain components in the
gaugino and higgsino directions, there is, strictly speaking, no way to assign 
the (SM) lepton number, hence (MSSM) R parity, to the list of GSSM superfields
free from ambiguity. In relation to that, we would go so far as to ask the readers 
to give up the preconception that R-parity violation is small, for the moment. 
Whether it is small is actually a delicate question, at least from the theoretical 
point of view. Among other things, there is the question of ``{ How small is 
small ?}". We will come back to all these below. Anyway, our formulation here
is generally valid. The only expressions inside the article with limited
validity are the perturbative diagonalization formulae of the mass matrices.
That perturbative regime is firmly based on phenomenology --- the fact that
neutrino masses are substantially smaller than the electroweak scale.

If the readers are willing, however, to forget totally about the notion of R parity 
for the moment and just follow our discussion here seeing the subject as what it is 
theoretically --- the GSSM, he or she will be in the best place to appreciate our 
discussion of the formulation part. In our opinion, that also set the best stage 
for looking at all the R-parity violation works --- something we will come back
to, in a Q \& A format, in the Appendix. 

In the section below we discuss the basic formulation and the issue of
parametrization. Then, we will focus on the so-called
single-VEV parametrization first explicitly advocated in Ref.\cite{ru1}.
Readers may first look at it as one possible parametrization of the GSSM. We will 
discuss some details of the model under the parametrization and illustrate 
its phenomenological merits. The formulation under the parametrization
will also provide a platform to address the various issues, to be discussed
below.  We elaborate on the content of the model using our formulation. We 
discuss in section~III and IV the fermion sectors and the scalar sectors, giving 
the explicit mass matrices, and perturbative diagonalization matrix elements in the 
phenomenologically interesting (small $\mu_i$) region. Such results are all
given here admitting the most generally nature of the parameters involved,
including possible nontrivial complex phases. Most of the results are taken from
previous works of the present author and various 
collaborators\cite{ru1,ru2,as5,as4,as6,as7,as9}.
Some of the expressions are however  not published before in the present 
most general form. In particular, most explicit results on the part of
the scalar mass matrices, used to some extent in Refs.\cite{as7,as9}, 
are not available elsewhere. In section V, we conclude this review with some 
remarks. In the Appendix section, we recapitulate on some of the important issues 
on the formulation aspect, and address other formulations and approaches
used in the literature, in a Q \& A format.

\section{Formulation of the GSSM} 
Let us start from the beginning and look carefully at the supersymmetrization 
of the SM. The gauge field sector is relatively trivial. 
In the matter field sector, all fermions and scalars have to
be promoted to chiral superfields containing both parts. It is 
straight forward for the quark doublets and singlets, and also for the leptonic
singlets. The leptonic doublets, however, have the same quantum number as
the Higgs doublet, $H_d$, that couples to the down-sector quarks. Nevertheless, 
one cannot simply get the Higgs, $H_d$, from the scalar partners of the leptonic 
doublets, $L$'s. Holomorphicity of the superpotential requires a separate 
superfield to contribute the Higgs which couples to the up-sector quarks.
$\!\!\!$\footnote{This is a constraint only when we insist on having the SM Yukawa
terms as the source of generation of up-sector quark masses. If one is ready 
to accept a loop level mass generation, a supersymmetric SM with only the
three leptonic doublets and singlets as the colorless chiral superfield is
certainly an interesting option, which has apparently not been explored.
However, with the not historically expected very large top mass confirmed,
such a model would have severe problem with top mass generation and hence
becomes very unappealing. 
} 
This $\hat{H}_u$ superfield then sure contributes an extra fermionic doublet, the 
higgsino, with nontrivial gauge anomalies. To cancel the latter, an extra fermionic 
doublet with the quantum number of $H_d$ or $L$ is needed. So, the result is that we 
need four superfields with that quantum number. As they are {\it a priori} 
indistinguishable, we label them by $\hat{L}_{\alpha}$ 
with the Greek subscript being an (extended) flavor index going from $0$ to $3$.

\subsection{The Superpotential and the Single-VEV Parametrization}
The most general renormalizable superpotential with the spectrum of minimal
superfields discussed above can be written as
\begin{equation}
W \!\! = \!\varepsilon_{ab}\left[ \mu_{\alpha}  \hat{H}_u^a \hat{L}_{\alpha}^b 
+ h_{ik}^u \hat{Q}_i^a   \hat{H}_{u}^b \hat{U}_k^{\scriptscriptstyle C}
+ \lambda_{\alpha jk}^{\!\prime}  \hat{L}_{\alpha}^a \hat{Q}_j^b
\hat{D}_k^{\scriptscriptstyle C} + 
\frac{1}{2}\, \lambda_{\alpha \beta k}  \hat{L}_{\alpha}^a  
 \hat{L}_{\beta}^b \hat{E}_k^{\scriptscriptstyle C} \right] + 
\frac{1}{2}\, \lambda_{ijk}^{\!\prime\prime}  
\hat{U}_i^{\scriptscriptstyle C} \hat{D}_j^{\scriptscriptstyle C}  
\hat{D}_k^{\scriptscriptstyle C} \;  ,
\end{equation}
where  $(a,b)$ are $SU(2)$ indices, $(i,j,k)$ are the usual family (flavor) 
indices (going from $1$ to $3$). We have explained the origin of the 4
$\hat{L}_{\alpha}$'s, with the $(\za, \zb)$ indices as extended flavor indices
going from $0$ to $3$. The rest of the superfield notations are obvious.
Note that $\lambda$ is antisymmetric in the first two indices, 
as required by the $SU(2)$  product rules, shown explicitly here with 
$\varepsilon_{\scriptscriptstyle 12} =-\varepsilon_{\scriptscriptstyle 21}=1$.
Similarly, $\lambda^{\!\prime\prime}$ is antisymmetric in the last two indices
from $SU(3)_{\scriptscriptstyle C}$, though color contents are not shown here.

Besides the superpotential, the Lagrangian contains the gauge interaction
part, including kinetic terms of the matter superfields, and a soft SUSY breaking
part. The former is trivial. The latter we will postpone till after we address
the question of choosing a specific parametrization for the theory.

First, it is important to note that after the supersymmetrization some of 
the superfields lose  the exact identities they have in relation to the 
physical particles as in the SM. The physical particles have to be mass 
eigenstates, which have to be worked out from the Lagrangian of the model.
Assuming electroweak symmetry breaking, we have now five (color-singlet) 
charged fermions, for instance. Involved in their masses are 1+4 admissible 
VEVs consistent with  the symmetry breaking, together with a SUSY
breaking gaugino mass. If one writes down naively the (tree-level) mass matrix,
the result is extremely complicated (see Ref.\cite{ru10} for an explicit 
illustration), with all the $\mu_{\scriptscriptstyle \za}$ and $\lambda_{\za\zb k}$ 
couplings involved.  Note that  the only definite experimental data we have 
here are the three physical lepton masses as the light eigenvalues, and the 
overall magnitude of the electroweak symmetry breaking VEVs. The task 
of analyzing the model seems to be formidable.

Recall that in the SM, the only three unit-charged fermions have a mass
matrix that is essentially diagonal. In another word, one can choose to write
the Lagrangian in the flavor basis corresponding to the physical charged
leptons $e, \mu,\,\mbox{and}\,\tau$. The leptonic Yukawa terms are then
by definition flavor diagonal, hence involving only 3 real parameters. One would 
like to achieve a similar simplification here. There is some problem though. We
have 4 ``leptonic doublets" $L_\za$'s, and, added to that, a gaugino from an
adjoint triplet all contributing to $e, \mu,\,\mbox{and}\,\tau$. As the superpotential
has to respect electroweak symmetry, mass eigenstate basis cannot be used here.
Choosing  flavor bases to write the Lagrangian is not just a matter of convenience.
Doing phenomenological studies without specifying a choice of flavor bases is
in fact ambiguous. A still better example is provided by thinking about doing SM 
quark physics with 18 complex Yukawa couplings, instead of the 10 real physical 
parameters, namely 6 quark masses and 4 real numbers needed to parametrize
the CKM matrix. As far as the SM itself is concerned, the extra 26 (real) 
parameters are simply redundant, and attempts to relate the full 36 parameters to
experimental data will be futile. There is simply no way to learn about the 36 real 
parameters of Yukawa couplings for the quarks in some generic flavor bases, so far 
as the SM is concerned.
$\!\!\!\!$\footnote{The full set of 36 parameters may be of interest
only when we want to model the origin of the flavor structure at a deeper 
level and hence higher energy scale\cite{hs}. Any meaningful attempts in the 
direction is likely to be possible only when we do have some good knowledge about
the phenomenological value of the SM parameters (in a specific basis). 
}
The best thing to do is to write the Lagrangian 
with a specific optimal set of parameters that also helps to simplify the 
analysis and make the physics more transparent. Again, this is exactly what we 
do with SM quark physics. For instance, one can choose to 
write the SM quark Yukawa couplings such that the down-quark Yukawa couplings 
are diagonal, while the up-quark Yukawa coupling matrix is a product of (the 
conjugate of) the CKM and the diagonal quark masses, and the leptonic Yukawa
couplings diagonal. 
$\!\!\!$\footnote{Here, what we are doing may be a bit unconventional when compared
with what is usually done in SM studies. However, our choice is well-motivated.
The down-sector quarks have extra couplings, the $\lambda^\prime$s, while
the up-sector is kept simple.}
Doing that has imposing no constraint or assumption onto the model. On the contrary, 
not fixing the flavor bases makes the connection between the parameters of the model 
and the phenomenological observables ambiguous.  

The choice of parametrization is not unique. However, a specific, consistent, 
choice has to be made before doing phenomenological studies --- before one uses
experimental data to constrain or pin down the value of any parameter. A
parametrization using generic flavor bases is ambiguous and redundant. 
In the case of the GSSM, the choice of flavor basis among the 4 $\hat{L}_\za$'s
is a particularly subtle issue, because of the fact that they are
superfields the scalar parts of which could bear VEVs. A parametrization
called the single-VEV parametrization (SVP) has been advocated by the author 
and collaborators since Ref.\cite{ru1}. The central idea is to pick a flavor
basis such that only one among the $\hat{L}_\za$'s, designated as $\hat{L}_0$,
bears a non-zero VEV. There is to say, the direction of the VEV, or the Higgs
field $H_d$, is singled out in the four dimensional vector space spanned by
the $\hat{L}_\za$'s. Explicitly, under the SVP, flavor bases are chosen such that : 
1/  $\langle \hat{L}_i \rangle \equiv 0$, which implies 
$\hat{L}_0 \equiv \hat{H}_d$;
2/  $y^{e}_{jk} (\equiv \lambda_{0jk} =-\lambda_{j0k}) 
=\frac{\sqrt{2}}{v_{\scriptscriptstyle 0}}\,{\rm diag}
\{m_{\scriptscriptstyle 1},
m_{\scriptscriptstyle 2},m_{\scriptscriptstyle 3}\}$;
3/ $y^{d}_{jk} (\equiv \lambda^{\!\prime}_{0jk}) 
= \frac{\sqrt{2}}{v_{\scriptscriptstyle 0}}\,{\rm diag}\{m_d,m_s,m_b\}$; 
4/ $y^{u}_{ik}=\frac{\sqrt{2}}{v_{\scriptscriptstyle u}}\,
V_{\!\mbox{\tiny CKM}}^{\!\scriptscriptstyle T}\; {\rm diag}\{m_u,m_c,m_t\}$, 
where $v_{\scriptscriptstyle 0}\equiv \sqrt{2} \, \langle \hat{L}_0 \rangle$
and $v_{\scriptscriptstyle u}\equiv \sqrt{2} \,
\langle \hat{H}_{u} \rangle$. 2/ to 4/ are more straight forward choices that
look just like the SM case, except that such choice can be consistently 
implemented in the GSSM case only because of choice 1/ --- an issue we will 
elaborate on below. The other important point to note is that the $m_i$'s above 
are, conceptually,  {\it not} the charged lepton masses. They are some unknown 
real parameters, though we will see below that they might turn out to be numerically
essentially the same as the charged lepton masses. The parametrization
given here still contains redundant complex phases among the couplings.
It is otherwise optimal. Removing the redundant phases is especially important
when we want to probe details of the CP violating  physics. However, for our discussion
here, we simply assume all the admissible nonzero couplings within the SVP
are generally complex.

Electroweak symmetry breaking is the sole source of chiral (SM) 
fermion masses, as well as the $LR$-mixings of their scalar superpartners. 
Hence, it is no surprise that the parametrization with the 
minimal number of nonzero VEVs gives the simplest structure for the 
(tree-level) mass matrices of both the fermions and the scalars. The adoption
of the SVP has, hence, a strong phenomenological advantage, which we will 
illustrate below. 

\subsection{The Soft SUSY Breaking Part}
Now we turn to the soft SUSY breaking part of the Lagrangian, details of which 
are too often overlooked. Following our notation above, the soft terms 
can be written as follow\cite{as5} :
\beqa
V_{\rm soft} &=& \epsilon_{\!\scriptscriptstyle ab} 
  B_{\za} \,  H_{u}^a \tilde{L}_\za^b +
\epsilon_{\!\scriptscriptstyle ab} \left[ \,
A^{\!\scriptscriptstyle U}_{ij} \, 
\tilde{Q}^a_i H_{u}^b \tilde{U}^{\scriptscriptstyle C}_j 
+ A^{\!\scriptscriptstyle D}_{ij} 
H_{d}^a \tilde{Q}^b_i \tilde{D}^{\scriptscriptstyle C}_j  
+ A^{\!\scriptscriptstyle E}_{ij} 
H_{d}^a \tilde{L}^b_i \tilde{E}^{\scriptscriptstyle C}_j   \,
\right] + {\rm h.c.}\nonumber \\
&+&
\epsilon_{\!\scriptscriptstyle ab} 
\left[ \,  A^{\!\scriptscriptstyle \lambda^\prime}_{ijk} 
\tilde{L}_i^a \tilde{Q}^b_j \tilde{D}^{\scriptscriptstyle C}_k  
+ \frac{1}{2}\, A^{\!\scriptscriptstyle \lambda}_{ijk} 
\tilde{L}_i^a \tilde{L}^b_j \tilde{E}^{\scriptscriptstyle C}_k  
\right] 
+ \frac{1}{2}\, A^{\!\scriptscriptstyle \lambda^{\prime\prime}}_{ijk} 
\tilde{U}^{\scriptscriptstyle C}_i  \tilde{D}^{\scriptscriptstyle C}_j  
\tilde{D}^{\scriptscriptstyle C}_k  + {\rm h.c.}
\nonumber \\
&+&
 \tilde{Q}^\dagger \tilde{m}_{\!\scriptscriptstyle {Q}}^2 \,\tilde{Q} 
+\tilde{U}^{\dagger} 
\tilde{m}_{\!\scriptscriptstyle {U}}^2 \, \tilde{U} 
+\tilde{D}^{\dagger} \tilde{m}_{\!\scriptscriptstyle {D}}^2 
\, \tilde{D} 
+ \tilde{L}^\dagger \tilde{m}_{\!\scriptscriptstyle {L}}^2  \tilde{L}  
  +\tilde{E}^{\dagger} \tilde{m}_{\!\scriptscriptstyle {E}}^2 
\, \tilde{E}
+ \tilde{m}_{\!\scriptscriptstyle H_{\!\scriptscriptstyle u}}^2 \,
|H_{u}|^2 
\nonumber \\
&& + \frac{M_{\!\scriptscriptstyle 1}}{2} \tilde{B}\tilde{B}
   + \frac{M_{\!\scriptscriptstyle 2}}{2} \tilde{W}\tilde{W}
   + \frac{M_{\!\scriptscriptstyle 3}}{2} \tilde{g}\tilde{g}
+ {\rm h.c.} \; ,
\label{soft}
\eeqa
where we have used ${H}_{d}$ in the place of the equivalent $\tilde{L}_0$ 
among the trilinear $A$-terms. Note that
$\tilde{L}^\dagger \tilde{m}_{\!\scriptscriptstyle \tilde{L}}^2  \tilde{L}$,
unlike the other soft mass terms, is given by a 
$4\times 4$ matrix. Comparing with the MSSM case,
$\tilde{m}_{\!\scriptscriptstyle {L}_{00}}^2$ corresponds to 
$\tilde{m}_{\!\scriptscriptstyle H_{\!\scriptscriptstyle d}}^2$ while 
$\tilde{m}_{\!\scriptscriptstyle {L}_{0k}}^2$'s give new mass mixings.
The other notations are obvious. The writing of the soft terms in the above 
form makes identification of the scalar mass terms straight forward. Recall
that only the doublets ${H}_{u}$ and ${H}_{d}$ bear VEVs. The $A$-terms
in the second line of Eq.(\ref{soft}) hence do not contribute to scalar masses.

For the sake of completeness, we include here the admissible nonholomorphic 
soft terms\cite{nhs,Ma}, to be given as
\beqa
V_{\rm soft}^{\scriptscriptstyle  \rm{NH}}  &=&  
  C^{\scriptscriptstyle U}_{ij} \, 
\tilde{Q}^a_i (H_{d}^a )^{\!*} \tilde{U}^{\scriptscriptstyle C}_j
+ C^{\scriptscriptstyle D}_{ij} 
(H_{u}^a)^{\!*} \tilde{Q}^a_i \tilde{D}^{\scriptscriptstyle C}_j  
+ C^{\scriptscriptstyle E}_{ij} 
(H_{u}^a)^{\!*} \tilde{L}^a_i \tilde{E}^{\scriptscriptstyle C}_j   \,
+ C^{\scriptscriptstyle H}_{k} (H_{u}^a)^{\!*} H_{d}^a 
 \tilde{E}^{\scriptscriptstyle C}_k
+ {\rm h.c.}\nonumber \\
&+&
  C^{\scriptscriptstyle L}_{ijk} 
(\tilde{L}_i^a)^{\!*} \tilde{Q}^a_j \tilde{U}^{\scriptscriptstyle C}_k    
+  C^{\scriptscriptstyle S}_{ijk} 
\tilde{U}^{\scriptscriptstyle C}_i  \tilde{E}^{\scriptscriptstyle C}_j  
(\tilde{D}^{\scriptscriptstyle C}_k)^{\!*}  
+ \frac{1}{2}\, \epsilon_{\!\scriptscriptstyle ab} 
C^{\scriptscriptstyle Q}_{ijk}
\tilde{Q}^a_i \tilde{Q}^b_j (\tilde{D}^{\scriptscriptstyle C}_k)^{\!*} 
+ {\rm h.c.}\; ,
\label{nhsoft}
\eeqa
where we have dropped bilinear terms which could be incorporated into 
$\tilde{m}_{\!\scriptscriptstyle {L}}^2$ above. Here again, only terms in the
first line could contribute to scalar masses.  

\subsection{Some Notation for the Component Fields}
Note that though the SVP enforces the identification of $\hat{L}_0$ as the one 
having ``Higgs" properties (of $\hat{H}_d$),  it still maintains couplings similar 
to those of the  $\hat{L}_i$'s. Put it in another way, the charged leptons in GSSM 
generally contain higgsino components, and the Higgs field may be partly the 
superpartners of the physical charged leptons.

We write the components of a $\hat{L}_\za$ fermion 
doublet as ${l}_{\!\scriptscriptstyle \za}^{\scriptscriptstyle 0}$
and ${l}_{\!\scriptscriptstyle \za}^{\!\!\mbox{ -}}$, and their scalar
partners as $\tilde{l}_{\!\scriptscriptstyle \za}^{\scriptscriptstyle 0}$
and $\tilde{l}_{\!\scriptscriptstyle \za}^{\!\!\mbox{ -}}$. Apart from being 
better motivated theoretically, the common notation helps to trace the 
flavor structure. However, we will also use notations of the form
${h}_{\!\scriptscriptstyle d}^{\star}$ ({\it i.e.} 
${h}_{\!\scriptscriptstyle d}^{\scriptscriptstyle 0}$ 
and ${h}_{\!\scriptscriptstyle d}^{\!\!\mbox{ -}}$)
$\tilde{h}_{\!\scriptscriptstyle d}^{\star}$, as alternative notations for
$\tilde{l}_{\!\scriptscriptstyle 0}^{\star}$ and
${l}_{\!\scriptscriptstyle 0}^{\star}$, in some places below. This is 
unambiguous under our formulation. We will also referred to the states
${h}_{\!\scriptscriptstyle d}^{\star}$ ($\equiv \tilde{l}_{\!\scriptscriptstyle 0}^{\star}$)
and $\tilde{h}_{\!\scriptscriptstyle d}^{\star}$ ($\equiv {l}_{\!\scriptscriptstyle 0}^{\star}$)
as Higgs and higgsino, respectively; while they are generally also included in the 
terms slepton and lepton.
 
In the left-handed lepton and slepton field notations
introduced above, we have dropped the commonly used $L$-subscript, for
simplicity. For the components of the three right-handed leptonic 
superfields, we use ${l}_i^{\scriptscriptstyle +}$ and
$\tilde{l}_i^{\scriptscriptstyle +}$, with again the $R$-subscript
dropped. The notation for the quark and squark fields will be standard,
with the $L$- and $R$-subscripts. A normal quark state, such as 
$d_{\!\scriptscriptstyle L_k}$, denotes a mass eigenstate, while a squark
state the supersymmetric partner of one. A quark or squark state with
a $\prime$ denotes one with the quark state being the $SU(2)$ partner 
of a mass eigenstate. For instance,  
$\tilde{u}^{\prime}_{\!\scriptscriptstyle L_3}$ is the up-type squark state
from $\hat{Q}_{\scriptscriptstyle 3}$ which contains the exact
left-handed $b$ quark according to our parametrization of the Lagrangian.
The scalar and fermion states of the up-sector Higgs doublet are denoted by 
${h}_{\!\scriptscriptstyle u}^{\!\scriptscriptstyle +}$ and
${h}_{\!\scriptscriptstyle u}^{\!\scriptscriptstyle 0}$, and
$\tilde{h}_{\!\scriptscriptstyle u}^{\!\scriptscriptstyle +}$ and
$\tilde{h}_{\!\scriptscriptstyle u}^{\!\scriptscriptstyle 0}$,
respectively. 

\subsection{Explicit Scalar-Fermion-Fermion Couplings}
We are now ready to spell out the couplings of the component fields. Of particular 
phenomenological interest are the scalar-fermion-fermion couplings. The gaugino 
couplings are, of course, standard. Coming from the gauge interaction parts, they are 
exactly the same as in MSSM. The couplings from the superpotential is, however,
much ridher in content.  As an explicit illustration of our notation and for easy 
reference, we list them here. Firstly, we give the corresponding couplings concerning 
the (color-singlet) charged and neutral fermion. from the superpotential. Compared
with that of MSSM, we have a modified higgsino part and a list of new terms from 
the trilinear couplings. We have 
\beqa
{\cal{L}}_{\chi}
&=&  y_{\!\scriptscriptstyle u_i} \,
V_{\!\mbox{\tiny CKM}}^{ij} \;
\tilde{h}_{\!\scriptscriptstyle u}^{\!\scriptscriptstyle +} \,
\left[ \;
\tilde{u}_{\!\scriptscriptstyle R_i}^c \,
d_{\!\scriptscriptstyle L_j} +
u_{\!\scriptscriptstyle R_i}^c \,
\tilde{d}_{\!\scriptscriptstyle L_j}  \,
\right] 
+ y_{\!\scriptscriptstyle d_i} \,
{l}_{\scriptscriptstyle 0}^{\!\!\mbox{ -}} \,
\left[ \;
\tilde{d}_{\!\scriptscriptstyle R_i}^c \,
u_{\!\scriptscriptstyle L_i}^{\prime}  +
d_{\!\scriptscriptstyle R_i}^c \,
\tilde{u}_{\!\scriptscriptstyle L_i}^{\prime}  \,
\right]  
 +  \lambda^{\!\prime}_{ijk} \;
{l}_i^{\!\!\mbox{ -}} \,
\left[ \;
\tilde{d}_{\!\scriptscriptstyle R_k}^c \,
u_{\!\scriptscriptstyle L_j}^{\prime}  +
d_{\!\scriptscriptstyle R_k}^c \,
\tilde{u}_{\!\scriptscriptstyle L_j}^{\prime}  \,
\right] 
\nonumber \\
&& - \;
y_{\!\scriptscriptstyle u_i} \,
\tilde{h}_{\!\scriptscriptstyle u}^{\!\scriptscriptstyle 0} \,
\left[ \;
\tilde{u}_{\!\scriptscriptstyle R_i}^c \,
u_{\!\scriptscriptstyle L_i} +
u_{\!\scriptscriptstyle R_i}^c \,
\tilde{u}_{\!\scriptscriptstyle L_i} \,
\right] 
- y_{\!\scriptscriptstyle d_i} \,
{l}_{\scriptscriptstyle 0}^{\scriptscriptstyle 0} \,
\left[ \;
\tilde{d}_{\!\scriptscriptstyle R_i}^c \,
d_{\!\scriptscriptstyle L_i}  +
d_{\!\scriptscriptstyle R_i}^c \,
\tilde{d}_{\!\scriptscriptstyle L_i} \,
\right]
- \lambda^{\!\prime}_{ijk} \;
{l}_i^{\scriptscriptstyle 0} \,
\left[ \;
\tilde{d}_{\!\scriptscriptstyle R_k}^c \,
d_{\!\scriptscriptstyle L_j}  +
d_{\!\scriptscriptstyle R_k}^c \,
\tilde{d}_{\!\scriptscriptstyle L_j}  \,
\right]
\nonumber \\ 
&& + \;
 y_{\!\scriptscriptstyle e_i} \,
\left[ \;
{l}_{\scriptscriptstyle 0}^{\!\!\mbox{ -}} \,
{l}_i^{\scriptscriptstyle +} \,
\tilde{l}_i^{\scriptscriptstyle 0}  \,
-
{l}_i^{\!\!\mbox{ -}}  \,
{l}_i^{\scriptscriptstyle +} \,
\tilde{l}_{\scriptscriptstyle 0}^{\scriptscriptstyle 0} \,
\right]
+ y_{\!\scriptscriptstyle e_i} \,
\left[ \;
{l}_{\scriptscriptstyle 0}^{\!\!\mbox{ -}} \,
{l}_i^{\scriptscriptstyle 0}  \,
\tilde{l}_i^{\scriptscriptstyle +} \,
-
{l}_i^{\!\!\mbox{ -}} \,
{l}_{\scriptscriptstyle 0}^{\scriptscriptstyle 0} \,
\tilde{l}_i^{\scriptscriptstyle +} \,
\right]
 + y_{\!\scriptscriptstyle e_i} \,
\left[ \;
{l}_i^{\scriptscriptstyle 0}  \,
{l}_i^{\scriptscriptstyle +} \,
\tilde{l}_{\scriptscriptstyle 0}^{\!\!\mbox{ -}} \,
-
{l}_{\scriptscriptstyle 0}^{\scriptscriptstyle 0} \,
{l}_i^{\scriptscriptstyle +} \,
\tilde{l}_i^{\!\!\mbox{ -}}  \,
\right] 
\nonumber \\
&& + \; 
 \lambda_{ijk} \;
{l}_i^{\!\!\mbox{ -}} \,
{l}_k^{\scriptscriptstyle +} \,
\tilde{l}_j^{\scriptscriptstyle 0}  \, 
+  \lambda_{ijk} \;
{l}_i^{\!\!\mbox{ -}} \,
{l}_j^{\scriptscriptstyle 0} \,
\tilde{l}_k^{\scriptscriptstyle +} 
- \lambda_{ijk} \;
{l}_i^{\scriptscriptstyle 0} \,
{l}_k^{\scriptscriptstyle +} \,
\tilde{l}_j^{\!\!\mbox{ -}}  \,
\quad + \quad \mbox{h.c.} \;\; ,
\label{Lc}\eeqa
where
\beq \label{yy}
y_{\!\scriptscriptstyle u_i} = 
\frac {g_{\scriptscriptstyle 2} \, m_{u_i} }
{\sqrt{2}\, M_{\!\scriptscriptstyle W} \,\sin\!\beta}
\;\; , \hspace{10mm}
y_{\!\scriptscriptstyle d_i} = 
\frac {g_{\scriptscriptstyle 2} \, m_{d_i} }
{\sqrt{2}\, M_{\!\scriptscriptstyle W} \,\cos\!\beta}
\;\; , \hspace{10mm}
y_{\!\scriptscriptstyle e_i} = 
\frac {g_{\scriptscriptstyle 2} \, m_{i} }
{\sqrt{2}\, M_{\!\scriptscriptstyle W} \,\cos\!\beta}
\;\; ,
\eeq
are the (diagonal) quark and charged lepton Yukawa couplings; and 
$\tan\!\beta =\frac{v_{\scriptscriptstyle u}}{v_{\scriptscriptstyle 0}}$
(see section II.D.below). Recall that $\lambda^{\!\prime}_{{\scriptscriptstyle 0}jk}$ 
corresponds to the down-Yukawa coupling matrix,  and 
$\lambda_{{\scriptscriptstyle 0}jk}$ corresponds to the charged  lepton Yukawa 
coupling matrix, both of which are diagonal under the SVP; in addition, we have 
$u_{\!\scriptscriptstyle L_i}^{\prime} = V_{\!\mbox{\tiny CKM}}^{\dag\, ij}\, 
u_{\!\scriptscriptstyle L_j}$ being $SU(2)$ partner of the mass eigenstate
$d_{\!\scriptscriptstyle L_i}$, and $\tilde{u}_{\!\scriptscriptstyle L_i}^{\prime}$
its scalar partner. We also use below 
$\tilde{d}_{\!\scriptscriptstyle L_i}^{\,\prime}$, which is, explicitly,
$V_{\!\mbox{\tiny CKM}}^{ij}\, \tilde{d}_{\!\scriptscriptstyle L_j}$.

There are some more scalar-fermion-fermion terms besides those given in 
${\cal{L}}_{\chi}$. These extra terms are slepton-quark-quark terms.  With the above 
explicit listed terms, however, it is straight forward to see what the extra 
terms are like.  They are given by
\beqa
{\cal{L}}_{qqs}
&=&   y_{\!\scriptscriptstyle u_i} \,
V_{\!\mbox{\tiny CKM}}^{ij} \;
{h}_{\!\scriptscriptstyle u}^{\!\scriptscriptstyle +} \,
{u}_{\!\scriptscriptstyle R_i}^c \,
d_{\!\scriptscriptstyle L_j} 
+ y_{\!\scriptscriptstyle d_i} \,
\tilde{l}_{\scriptscriptstyle 0}^{\!\!\mbox{ -}} \,
{d}_{\!\scriptscriptstyle R_i}^c \,
u_{\!\scriptscriptstyle L_i}^{\prime} 
 +  \lambda^{\!\prime}_{ijk} \;
\tilde{l}_i^{\!\!\mbox{ -}} \,
{d}_{\!\scriptscriptstyle R_k}^c \,
\nonumber \\
&& - \;
y_{\!\scriptscriptstyle u_i} \,
{h}_{\!\scriptscriptstyle u}^{\!\scriptscriptstyle 0} \,
{u}_{\!\scriptscriptstyle R_i}^c \,
u_{\!\scriptscriptstyle L_i} 
- y_{\!\scriptscriptstyle d_i} \,
\tilde{l}_{\scriptscriptstyle 0}^{\scriptscriptstyle 0} \,
{d}_{\!\scriptscriptstyle R_i}^c \,
d_{\!\scriptscriptstyle L_i} 
 - \lambda^{\!\prime}_{ijk} \;
\tilde{l}_i^{\scriptscriptstyle 0} \,
{d}_{\!\scriptscriptstyle R_k}^c \,
d_{\!\scriptscriptstyle L_j} 
\quad + \quad \mbox{h.c.} \;\; .
\label{qqs}\eeqa

In both of the above expressions for ${\cal{L}}_{\chi}$, the terms present in MSSM  
can be easily identified, with the replacement of the ${l}_{\scriptscriptstyle 0}^{\star}$
and $\tilde{l}_{\scriptscriptstyle 0}^{\star}$ states by the more familiar notation of 
$\tilde{h}_{\scriptscriptstyle d}^{\star}$ and ${h}_{\scriptscriptstyle d}^{\star}$.
The nice feature, obtained without approximation, is a consequence of the SVP. The 
simple structure of the trilinear coupling contributions to the $d$-quark and charged 
lepton masses, is what make the analysis simple and easy to handle. We want to 
emphasize that the above expressions are exact tree-level results without hidden 
assumptions behind its validity. The only point of caution here is that the 
$l_i^{\star}$ states are not exactly the charged leptons and neutrinos. We will come
to the mass matrices for the fermions in the next section.

\subsection{Notes on the Scalar Potential}
As the SVP, conceptually, involves identifying the direction of the VEV
among the $\hat{L}_\za$'s, we will take a look at the scalar potential here
and see what this really means. The discussion also serves to answer queries
on whether this can be consistently done --- a question that causes some 
confusion. The major part of the results here is first presented in the
appendix of Ref.\cite{as5}.

In terms of the five, plausibly electroweak symmetry breaking, neutral 
scalars fields $\phi_n$, the generic (tree-level) scalar potential, as 
constrained by SUSY, can be written as :
\beqa
V_{\!\mbox{\tiny EW}\!\!\!\!\!\!\!\!//} &=& Y_{n} \left|\phi_n\right|^4 
+ X_{mn}  \left|\phi_m\right|^2 \left|\phi_n\right|^2 
+  \hat{m}^2_{n} \left|\phi_n\right|^2 
\nonumber \\
 && - ( \hat{m}^2_{\scriptscriptstyle m\!n} e^{i\theta\!_{m\!n}}
 \phi_m^{\dag} \phi_n + \mbox{h.c.}) \qquad\qquad\qquad (m < n) \; .
\eeqa
Here, we count the $\phi_n$'s from $-1$ to $3$ and identify a $\phi_\za$
(recall $\za=0$ to $3$) as 
$\tilde{l}_{\!\scriptscriptstyle \za}^{\scriptscriptstyle 0}$ 
and $\phi_{\scriptscriptstyle \mbox{-}\!1}$ as 
${h}_{\!\scriptscriptstyle u}^{\!\scriptscriptstyle 0}$.
Parameters (all real) in the above expression for 
$V_{\!\mbox{\tiny EW}\!\!\!\!\!\!\!\!//}$ are then given by
\beqa
 \hat{m}^2_{\za} &=& \tilde{m}^2_{\!{\scriptscriptstyle L\!_{\za\za}}}
 + \left|\mu_\za\right|^2 \; , \nonumber \\
 \hat{m}^2_{\scriptscriptstyle \mbox{-}\!1} &=& 
 \tilde{m}_{\!\scriptscriptstyle H\!_u}^2  + \mu_\za^{*} \mu_\za \; ,
\nonumber \\
\hat{m}^2_{\!\za\!\zb}\, e^{i\theta\!_{\za\!\zb}} &=& 
- \tilde{m}^2_{\!{\scriptscriptstyle L}\!_{ \za\!\zb}}
-  \mu_\za^{*} \mu_\zb \quad\quad  \mbox{(no sum)}\; ,	\nonumber \\
\hat{m}^2_{\!{\scriptscriptstyle \mbox{-}\!1\!\za}}\, 
e^{i\theta\!_{\scriptscriptstyle \mbox{-}\!1\!\za}} &=& B_\za 
\qquad\qquad\qquad\quad  \mbox{(no sum)}\; , \nonumber \\
 Y_{n} &=& 	\frac{1}{8}(g^2_{\!\scriptscriptstyle 1} 
 + g_{\!\scriptscriptstyle 2}^2)\; ,	\nonumber \\
 X_{{\scriptscriptstyle \mbox{-}\!1\!\za}}  &=& - 
\frac{1}{4}(g_{\!\scriptscriptstyle 1}^2 
 + g_{\!\scriptscriptstyle 2}^2) = -
 X_{{\scriptscriptstyle \za\!\zb}} \;.
\eeqa
Under the SVP, we write the VEVs as follow :
\beqa 
v_{\scriptscriptstyle \mbox{-}\!1}\,  (\equiv \sqrt{2}\,
\lla \phi_{\scriptscriptstyle \mbox{-}\!1} \rra)
& =& v_{\!\scriptscriptstyle u} \; , \nonumber \\
v_{\!\scriptscriptstyle 0} \, (\equiv \sqrt{2}\,
\lla \phi_{\scriptscriptstyle 0} \rra)
& =& v_{\!\scriptscriptstyle d}\, e^{i\theta\!_v} \; , \nonumber \\
v_{\scriptscriptstyle i} \, (\equiv \sqrt{2}\,
\lla \phi_i \rra)& =& 0 \; ,
\eeqa 
where we have put in a complex phase in the VEV $v_{\!\scriptscriptstyle 0}$,
for generality. 

The equations from the vanishing derivatives of 
$V_{\!\mbox{\tiny EW}\!\!\!\!\!\!\!\!//}$  along
$\phi_{\scriptscriptstyle \mbox{-}\!1}$ and $\phi_{\scriptscriptstyle 0}$
give
\beqa
{ \left[ \frac{1}{8}(g_{\!\scriptscriptstyle 1}^2  + g_{\!\scriptscriptstyle 2}^2) 
(v_{\!\scriptscriptstyle u}^2 -v_{\!\scriptscriptstyle d}^2) + 
\hat{m}^2_{\scriptscriptstyle \mbox{-}\!1} \right]} 
\, v_{\!\scriptscriptstyle u}
&=&
B_0 \, v_{\!\scriptscriptstyle d} \, e^{i\theta\!_v} \; ,
\nonumber \\
{ \left[ \frac{1}{8}(g_{\!\scriptscriptstyle 1}^2  + g_{\!\scriptscriptstyle 2}^2) 
(v_{\!\scriptscriptstyle d}^2 -v_{\!\scriptscriptstyle u}^2) + 
\hat{m}^2_{\scriptscriptstyle 0} \right] }
\, v_{\!\scriptscriptstyle d}
&=&
B_0 \, v_{\!\scriptscriptstyle u} \, e^{i\theta\!_v} \; .
\eeqa
Hence, $B_0 \, e^{i\theta\!_v}$ is real. In fact, the part of 
$V_{\!\mbox{\tiny EW}\!\!\!\!\!\!\!\!//}$  that
is relevant to obtaining the tadpole equations is no different from
that of MSSM apart from the fact that 
$\tilde{m}_{\!\scriptscriptstyle H\!_u}^2$ and 
$\tilde{m}_{\!\scriptscriptstyle H\!_d}^2$ of the latter are replaced by
$\hat{m}^2_{\scriptscriptstyle \mbox{-}\!1}$ and
$\hat{m}^2_{\scriptscriptstyle 0}$ respectively. As in MSSM, the $B_0$
parameter can be taken as real. The conclusion here
is therefore that $\theta\!_v$ vanishes, or all VEVs are real, despite
the existence of complex parameters in the scalar potential. The above tadpole
equations could then be written as 
\beqa
B_0 \, \cot\!\zb &=&
 \left[ \tilde{m}^2_{\!{\scriptscriptstyle H}_{\!u}}
+ \mu_{\scriptscriptstyle \za}^* \, \mu_{\scriptscriptstyle \za}
+\frac{1}{8}(g_{\!\scriptscriptstyle 1}^2  + g_{\!\scriptscriptstyle 2}^2) 
(v_{\!\scriptscriptstyle u}^2 -v_{\!\scriptscriptstyle d}^2) \right] \; ,
\nonumber \\
B_0 \, \tan\!\zb &=&
 \left[ \tilde{m}^2_{\!{\scriptscriptstyle L}_{\!{\scriptscriptstyle 00}}}
+ |\mu_{\scriptscriptstyle 0}|^2
+ \frac{1}{8}(g_{\!\scriptscriptstyle 1}^2  + g_{\!\scriptscriptstyle 2}^2) 
(v_{\!\scriptscriptstyle d}^2 -v_{\!\scriptscriptstyle u}^2)  \right]  \; .
\label{tp}
\eeqa
Results from the other tadpole equations, in a $\phi_i$ direction, are quite
simple. They can be written as complex equations of the form 
\begin{equation}
 \hat{m}^2\!\!_{{\scriptscriptstyle \mbox{-}\!1}i}\, 
e^{i\theta\!_{{\scriptscriptstyle \mbox{-}\!1}i}} \tan\!\beta
= -  e^{i\theta\!_v}
 \hat{m}^2\!\!_{{\scriptscriptstyle 0}i}\, 
e^{i\theta\!_{{\scriptscriptstyle 0}i}} \; ,
\end{equation}
which is equivalent to 
\begin{equation} \label{tp3}
B_i \, \tan\!\beta 
=     \tilde{m}^2_{\!{\scriptscriptstyle L}_{\!{\scriptscriptstyle 0}i} }
+ \mu_{\scriptscriptstyle 0}^{*}\,\mu_i \; ,
\end{equation} 
where we have used $v_{\scriptscriptstyle u}=v\sin\!\beta$ and
$v_{\scriptscriptstyle d}=v\cos\!\beta$. Note that our $\tan\!\beta$ has 
the same physical meaning as that in the MSSM case. For
instance, $\tan\!\beta$, together with the corresponding Yukawa coupling 
ratio, gives the mass ratio between the top and the bottom quarks.

The three complex equations for the $B_i$'s reflect the redundancy of 
parameters in a generic $\hat{L}_\za$ flavor basis. The equations also 
suggest that the $B_i$'s are expected to be suppressed, with respect to
the  $B_{\scriptscriptstyle 0}$, as the $\mu_i$'s are,
with respect to $\mu_{\scriptscriptstyle 0}$. They give consistence
relationships among the involved parameters (under the SVP) that
should not be overlooked. The 
$ \tilde{m}^2_{\!{\scriptscriptstyle L}_{\!{\scriptscriptstyle 0}i} }$ 
parameters in particular are missing in some of the relevant discussions
in the literature. From a different perspective, one may tend to think that
the parameters are similar to the $ \tilde{m}^2_{\!{\scriptscriptstyle L}_{\!ij} }$ 
parameters linked to soft flavor mixings. However, fixing 
 $ \tilde{m}^2_{\!{\scriptscriptstyle L}_{\!{\scriptscriptstyle 0}i} }$ 
in Eq.(\ref{tp3}) leads to definite relations between a $B_i$ and a $\mu_i$
term, which may not be satisfied {\it a priori}. The parameters $B_i$, $\mu_i$,
and $ \tilde{m}^2_{\!{\scriptscriptstyle L}_{\!{\scriptscriptstyle 0}i} }$ 
are not independent free parameters, because of the fact that freely chosen values 
of the set of parameters in a top-down approach, in general, do not land the model
automatically into the single-VEV basis\cite{align}. One can think about then
performing a flavor basis rotation to recast the model into the SVP
framework. The basis rotation would necessarily produce rotated $B_i$, $\mu_i$, 
and $ \tilde{m}^2_{\!{\scriptscriptstyle L}_{\!{\scriptscriptstyle 0}i} }$ 
parameters satisfying the tadpole equations. Nonzero values of 
$ \tilde{m}^2_{\!{\scriptscriptstyle L}_{\!{\scriptscriptstyle 0}i} }$ 
would be generated, for instance, even if one starts by choosing them to
be zero through whatever SUSY breaking scenario consideration. Whether
a more fundamental theory such as a specific SUSY breaking theory
with a particular prediction on the flavor structure of the GSSM 
would be automatically compatible with  the single-VEV basis is a
very interesting problem to be explored.
 
\section{The (Color-singlet) Fermions }
In this section, we start to look at the (tree-level) mass matrices of the matter
fields. We discuss the fermions here, and the scalars in the next section. We 
are quite ignorant about the scalar sector, knowing only that if the particles
exist at all, they have to be relatively heavy. For the fermions, we have some 
light particles observed, namely the charged leptons $e,\mu,$ and $\tau$ and the
neutrinos. In the GSSM framework, however, the neutrinos may not be
exact $SU(2)$ partners of the charged leptons. The known charged leptons
physical masses correspond only to mass eigenvalues of a big matrix 
incorporating also two heavy particles called charginos. For the neutral fermions,
we know that there have to be three very light states, corresponding to the
(physical neutrinos), and four heavy states called neutralinos. There is only some
information on likely oscillations among the neutrinos. Nevertheless, using the 
popular notion that all neutrinos have masses at or below the sub-eV scale gives 
quite stringent constraints on parameters such as the $\mu_i$'s. This is the 
scenario that we will mainly focus on. Within such a scenario, we will discuss 
perturbative diagonalizations of the mass matrices explicitly. The range of 
validity of our diagonalization results will be self-illustrative. Such results are 
useful in various phenomenological studies. For the generic scenario in 
which our analytical perturbative diagonalization procedures fail, the 
diagonalizations and, hence, the relationship between the physical states and
the states used to write the Lagrangian could only be extracted through numerical 
procedures. While the numerical approach could always be used, our analytical
diagonalization results are very useful for a clear understanding of the various
phenomenological features (see Refs.\cite{ru2,as5,as4,as6,as7,as9} for illustrations).

The SVP gives simple and straighforward tree-level mass matrices for the quarks 
in  exactly the same form as in the SM, without any approximation or simplifying 
assumption. Hence, we do not discuss the quark mass expressions further.
We would like to emphasize, however, that this is not the case if one works
in an alternative flavor basis. If the $\hat{L}_i$ VEVs are not zero, they
contribute to down-sector quark masses through the $\lambda^{\!\prime}_{ijk}$
couplings and the Lagrangian or superpotential can no longer be within
in the mass eigenstate basis of the quarks. Given the small masses of the
down and strange quarks, one may have to be particularly cautious on neglecting
such new contributions.

\subsection{Charged Fermions}
Under the SVP, the (color-singlet) charged fermion mass 
matrix is given by the simple form :
\beq \label{mc}
{\mathcal{M}_{\scriptscriptstyle C}} =
 \left(
{\begin{array}{ccccc}
{M_{\!\scriptscriptstyle 2}} &  {\sqrt 2} \, M_{\!\scriptscriptstyle W} \cos\!\beta
& 0 & 0 & 0 \\
  {\sqrt 2} \, M_{\!\scriptscriptstyle W} \sin\!\beta
&  {{ \mu}_{\scriptscriptstyle 0}} & {{ \mu}_{\scriptscriptstyle 1}} 
&  {{ \mu}_{\scriptscriptstyle 2}} & {{ \mu}_{\scriptscriptstyle 3}} \\
0 &  0 & \;\; {{m}_{\scriptscriptstyle 1}} \;\;  & 0 & 0 \\
0 & 0 & 0 & \;\; {{m}_{\scriptscriptstyle 2}} \;\; & 0 \\
0 & 0 & 0 & 0 & \;\; {{m}_{\scriptscriptstyle 3}} \;\; 
\end{array}}
\right)  \; ,
\eeq
with explicit bases for right-handed and left-handed states given by
$(-i\tilde{W}^{\scriptscriptstyle +},
\tilde{h}_{\!\scriptscriptstyle u}^{\!\scriptscriptstyle +},
{l}_{\scriptscriptstyle 1}^{\!\scriptscriptstyle +},
{l}_{\scriptscriptstyle 2}^{\!\scriptscriptstyle +},
{l}_{\scriptscriptstyle 3}^{\!\scriptscriptstyle +}\,)$
and
$(-i\tilde{W}^{\!\!\mbox{ -}},
{l}_{\scriptscriptstyle 0}^{\!\!\mbox{ -}},
{l}_{\scriptscriptstyle 1}^{\!\!\mbox{ -}},
{l}_{\scriptscriptstyle 2}^{\!\!\mbox{ -}},
{l}_{\scriptscriptstyle 3}^{\!\!\mbox{ -}}\,)$,
respectively. Here, we allow ${M_{\!\scriptscriptstyle 2}}$ and all four
$\mu_{\scriptscriptstyle \za}$ parameters to be complex, though the $m_i$'s
are mostly restricted to be real, for reason that would become clear below.
Obviously, each  $\mu_i$ parameter here characterizes 
directly the deviation of the ${l}_i^{\!\!\mbox{ -}}$ from the corresponding 
physical charged lepton  ($\ell_i = e$, $\mu$, and $\tau$), {\it i.e.} light 
mass eigenstates. For any set of other parameter inputs, the ${m}_i$'s can 
then be determined, through a simple numerical procedure, to guarantee that 
the correct mass eigenvalues of  $m_e$, $m_\mu$, and $m_\tau$  are obtained 
--- an issue first addressed and solved in Refs.\cite{ru1,ru2}. The latter issue
is especially important when $\mu_i$'s not substantially smaller than
${ \mu}_{\scriptscriptstyle 0}$ are considered. Such an odd scenario is
not definitely ruled out\cite{ru2}. 
However, for the more popular of small-$\mu_i$ scenario,
we have ${l}_i^{\!\!\mbox{ -}} \approx \ell_i^{\!\!\mbox{ -}}$, and deviations
of the ${l}_i^{\scriptscriptstyle +}$'s from  mass eigenstates 
$\ell_i^{\scriptscriptstyle +}$'s and $m_i$'s from the (real) $\ell_i$ 
masses are very negligible.

We introduce unitary matrices {\boldmath $V$} and {\boldmath $U$} diagonalizing 
the $R$- and $L$-handed states with
\beq
 \label{umcv}
 \mbox{\boldmath $V$}^\dag {\mathcal{M}_{\scriptscriptstyle C}} \,
\mbox{\boldmath $U$} = \mbox{diag} 
\{ {M}_{\!\scriptscriptstyle \chi^{\mbox{-}}_n} \} \equiv 
\mbox{diag} 
\{ {M}_{c {\scriptscriptstyle 1}}, {M}_{c {\scriptscriptstyle 2}},
m_e, m_\mu, m_\tau \}\; .
\eeq			
Here, the mass eigenvalues ${M}_{\!\scriptscriptstyle \chi^{\mbox{-}}_n}$ 
with $n=1$ and $2$, {\it i.e.}  
${M}_{c {\scriptscriptstyle 1}}$ and ${M}_{c {\scriptscriptstyle 2}}$, 
are the chargino masses. Note that notation here is different from those given in 
Refs.\cite{ru1,ru2}, and many others in the literature. More explicitly, we have 
$R$- and $L$-handed mass eigenstates given by
$(\, \chi_{{\!\scriptscriptstyle +}n} \,)= \mbox{\boldmath $V$}^{\scriptscriptstyle T} \,
[ -i\tilde{W}^{\scriptscriptstyle +},
\tilde{h}_{\!\scriptscriptstyle u}^{\!\scriptscriptstyle +},
{l}_{\scriptscriptstyle 1}^{\!\scriptscriptstyle +},
{l}_{\scriptscriptstyle 2}^{\!\scriptscriptstyle +},
{l}_{\scriptscriptstyle 3}^{\!\scriptscriptstyle +}\,]^{\scriptscriptstyle T}$
and
$(\, \chi_{\!\!\!\mbox{ -}n} \,)= \mbox{\boldmath $U$}^{\dag} \,
[-i\tilde{W}^{\!\!\mbox{ -}},
{l}_{\scriptscriptstyle 0}^{\!\!\mbox{ -}},
{l}_{\scriptscriptstyle 1}^{\!\!\mbox{ -}},
{l}_{\scriptscriptstyle 2}^{\!\!\mbox{ -}},
{l}_{\scriptscriptstyle 3}^{\!\!\mbox{ -}}\,]^{\scriptscriptstyle T}$; 
which form the five Dirac fermions 
${\chi}_n^{\!\!\mbox{ -}} = \left( \begin{array}{c}
\chi_{\!\!\!\mbox{ -}n} \\
\chi_{{\!\scriptscriptstyle +}n}^{\dag} 
\end{array} \right)$. 
Consider further
\beq
R_{\!\scriptscriptstyle R}^{\dagger} 
  \left( \begin{array}{cc}
  M_{\!\scriptscriptstyle 2}  & {\sqrt 2} \, M_{\!\scriptscriptstyle W} \cos\!\beta \\
 {\sqrt 2} \, M_{\!\scriptscriptstyle W} \sin\!\beta
& {\mu}_{\scriptscriptstyle 0}
  \end{array} \right) R_{\!\scriptscriptstyle L}
  \; = \; {\rm diag} 
  \{ M_{c{\scriptscriptstyle 1}}^{o}, M_{c{\scriptscriptstyle 2}}^{o} \} \;\;\; 
\label{Mcidef}
\eeq
with $M_{c{\scriptscriptstyle 1}}^{o}$ and $M_{c{\scriptscriptstyle 2}}^{o}$ 
being the chargino masses in the ${\mu}_i = 0$ limit. One can then write
the diagonalizing matrices in the block form
\begin{eqnarray} \label{VU}
\mbox{\boldmath $V$} =
  \left( \begin{array}{cc}
  R_{{\!\scriptscriptstyle R}}
  & -R_{{\!\scriptscriptstyle R}} \,V^{\dagger} \\
  V             & I_{\!\scriptscriptstyle 3 \times 3}                   
  \end{array} \right)  
\qquad \mbox{and} \qquad
\mbox{\boldmath $U$} =
  \left( \begin{array}{cc}
  R_{{\!\scriptscriptstyle L}}
  & -R_{{\!\scriptscriptstyle L}}\, U^{\dagger} \\
  U              & I_{\!\scriptscriptstyle 3 \times 3}                   
  \end{array} \right) \; .
\end{eqnarray}
Elements in the $R_{{\!\scriptscriptstyle R}}$ and $R_{{\!\scriptscriptstyle L}}$
matrices are all expected to be of order 1.
For ${\mu}_i \ll M_{c{\scriptscriptstyle a}}^{o}$ ($a=1$ and $2$),
a block perturbative diagonalization can be performed directly on the matrix 
${\mathcal{M}_{\scriptscriptstyle C}}$ to obtain the following simple result :
\beqa
\mbox{\boldmath $U$}_{\!(i+2)1} & \simeq& 
\frac{  \mu_i^* }{M_{c{\scriptscriptstyle 1}}} \,
R_{{\!\scriptscriptstyle R}_{21}}\; ,
\nonumber \\
\mbox{\boldmath $U$}_{\!(i+2)2} & \simeq& 
\frac{ \mu_i^*  }{M_{c{\scriptscriptstyle 2}}} \,
R_{{\!\scriptscriptstyle R}_{22}}\; ,
\nonumber \\
\mbox{\boldmath $V$}_{\!\!(i+2)a} & \simeq& 
\frac{m_i }{M_{c_{\scriptscriptstyle a}}} \,
\mbox{\boldmath $U$}_{\!(i+2)a}
\qquad\qquad (a=1 \,\mbox{and } 2)\; .
\label{VUele}
\eeqa
(Note : the $m_i$'s are assumed to be real here.)
The above expressions give the strength of the given matrix elements in the 
$U$ and $V$ blocks of Eq.(\ref{VU}). We note that the $L$-handed mixings are roughly 
measured by the ratio of a $\mu_i$ to the chargino mass scale, while $R$-handed 
mixings are further suppressed by a charged lepton to chargino mass ratio. 
The $\mbox{\boldmath $U$}_{\!(i+2)a}$ elements as given
above show no obvious dependence on $\tan\!\beta$, though some nontrivial
dependence is expected through the $R_{{\!\scriptscriptstyle R}_{2a}}$ elements.
An exact numerical study also confirms a weak sensitivity on the
$\tan\!\beta$ value (see Ref.\cite{as7} for example). 

On the other hand, we have, from Eqs.(\ref{VU}) and (\ref{VUele}), 
\begin{eqnarray}
\mbox{\boldmath $U$}_{\!\!a(i+2)} & \simeq &
- \mu_i \cdot \left[ R_{{\!\scriptscriptstyle L}} \, 
\left({\rm diag} 
  \{ M_{c{\scriptscriptstyle 1}}^{o}, M_{c{\scriptscriptstyle 2}}^{o} \}\right)^{-1} \, R_{{\!\scriptscriptstyle R}}^{\dag} \right]_{a2} \; ,
\nonumber \\
\mbox{\boldmath $V$}_{\!\!a(i+2)} & \simeq &
- \mu_i \cdot \left[ R_{{\!\scriptscriptstyle R}} \, 
\left( {\rm diag} 
  \{ M_{c{\scriptscriptstyle 1}}^{o}, M_{c{\scriptscriptstyle 2}}^{o} \}\right)^{-2} \, R_{{\!\scriptscriptstyle R}}^{\dag} \right]_{a2} \; ;
\eeqa
giving the result
\beqa
\mbox{\boldmath $U$}_{\!1(i+2)} & \simeq &
\frac{ \mu_i \, \sqrt{2}\, M_{\!\scriptscriptstyle W} \cos\!\beta}
{M_{\!\scriptscriptstyle 0}^2} \; ,
\nonumber \\
\mbox{\boldmath $U$}_{\!2(i+2)} & \simeq &
 -\frac{\mu_i \, M_{\!\scriptscriptstyle 2}}{M_{\!\scriptscriptstyle 0}^2} \; ,
\nonumber \\
\mbox{\boldmath $V$}_{\!\!1(i+2)} & \simeq & \mu_i \, m_i \;
 \frac{\sqrt{2}\, \, M_{\!\scriptscriptstyle W} ( M_{\!\scriptscriptstyle 2}^*\,\sin\!\beta
   + {\mu}_{\scriptscriptstyle 0}^* \,\cos\!\beta )}{|M_{\!\scriptscriptstyle 0}|^4} \; ,
\nonumber \\
\mbox{\boldmath $V$}_{\!\!2(i+2)} & \simeq & - \mu_i \, m_i \;
          \frac{( |M_{\!\scriptscriptstyle 2}|^2
  + 2 \, M_{\!\scriptscriptstyle W}^2 \, {\cos}^2\!\!{\beta} )}
{|M_{\!\scriptscriptstyle 0}|^4} \; ,
    \end{eqnarray}
where
\[
M_{\!\scriptscriptstyle 0}^2 \; \equiv \;
{\mu}_{\scriptscriptstyle 0} \, M_{\!\scriptscriptstyle 2} - 
M_{\!\scriptscriptstyle W}^2 \, \sin\! 2\beta  \; ,
\]
with magnitude given by $M_{c{\scriptscriptstyle 1}}^{o} \, M_{c{\scriptscriptstyle 2}}^{o}
\simeq M_{c{\scriptscriptstyle 1}} \, M_{c{\scriptscriptstyle 2}}$.
These matrix elements correspond to those given in Ref.\cite{ru1,ru2}, where 
only real parameters are taken. The crucial $\cos\!\beta$ dependence of
the non-standard $Z^0$-boson couplings of the physical charged leptons 
($\ell_i \equiv \chi_{\scriptscriptstyle i+2}$) through
$\mbox{\boldmath $U$}_{\!1(i+2)}$ is emphasized in the latter study. 
The different  $\tan\!\beta$ dependence between $\mbox{\boldmath $U$}_{\!(i+2)a}$ 
and $\mbox{\boldmath $U$}_{\!\!a(i+2)}$ is hence very important.

\subsection{Neutral Fermions}
The $7\times 7$ Majorana mass matrix for the neutral fermion can be written as
\small\begin{equation}
\label{mn}
\cal{M_{\scriptscriptstyle N}} = 
\left (\begin{array}{ccccccc}
M_{\!\scriptscriptstyle 1} & 0 
& M_{\!\scriptscriptstyle Z} \sin\!\theta_{\!\scriptscriptstyle W}  \sin\!\zb
& - M_{\!\scriptscriptstyle Z} \sin\!\theta_{\!\scriptscriptstyle W}  \cos\!\zb
 & 0 & 0 & 0  \\
0   & M_{\!\scriptscriptstyle 2} 
& -  M_{\!\scriptscriptstyle Z} \cos\!\theta_{\!\scriptscriptstyle W}  \sin\!\zb
&  M_{\!\scriptscriptstyle Z} \cos\!\theta_{\!\scriptscriptstyle W}  \cos\!\zb
 & 0 & 0 & 0  \\
M_{\!\scriptscriptstyle Z} \sin\!\theta_{\!\scriptscriptstyle W}  \sin\!\zb
 & - M_{\!\scriptscriptstyle Z} \cos\!\theta_{\!\scriptscriptstyle W}  \sin\!\zb
& 0  & -\mu_{\scriptscriptstyle 0}  & -\mu_{\scriptscriptstyle 1} 
& -\mu_{\scriptscriptstyle 2} & -\mu_{\scriptscriptstyle 3} \\
- M_{\!\scriptscriptstyle Z} \sin\!\theta_{\!\scriptscriptstyle W}  \cos\!\zb
& M_{\!\scriptscriptstyle Z} \cos\!\theta_{\!\scriptscriptstyle W}  \cos\!\zb
 & -\mu_{\scriptscriptstyle 0} &   0 & 0 & 0 & 0\\
  0  & 0  & -{\mu}_{\scriptscriptstyle 1}   & 0 & 0 & 0 & 0 \\
  0  & 0  & -{\mu}_{\scriptscriptstyle 2}   & 0 & 0 & 0 & 0 \\
  0  & 0  & -{\mu}_{\scriptscriptstyle 3}   & 0 & 0 & 0 & 0 
  \end{array} \right) \; ,
\eeq \normalsize
with explicit basis  $(-i\tilde{B}, -i\tilde{W}, 
\tilde{h}_{\!\scriptscriptstyle u}^{\!\scriptscriptstyle 0^C}\,, 
\tilde{h}_{\!\scriptscriptstyle d}^{\!\scriptscriptstyle 0}\,, 
{l}_{\scriptscriptstyle 1}^{\scriptscriptstyle 0}\,,
{l}_{\scriptscriptstyle 2}^{\scriptscriptstyle 0}\,,
{l}_{\scriptscriptstyle 3}^{\scriptscriptstyle 0}\,) $. 
Note that $\tilde{h}_{\!\scriptscriptstyle d}^{\!\scriptscriptstyle 0}
\equiv {l}_{\scriptscriptstyle 0}^{\scriptscriptstyle 0}$,
while $\tilde{h}_{\!\scriptscriptstyle u}^{\!\scriptscriptstyle 0^C}$
is the charge conjugate of the higgsino
$\tilde{h}_{\!\scriptscriptstyle u}^{\!\scriptscriptstyle 0}$ ; and,
from the above discussion of the charged fermions, 
we have, for small $\mu_i$'s,
$(\, {l}_{\scriptscriptstyle 1}^{\scriptscriptstyle 0},
{l}_{\scriptscriptstyle 2}^{\scriptscriptstyle 0},
{l}_{\scriptscriptstyle 3}^{\scriptscriptstyle 0} \, ) 
\approx (\nu_{\scriptscriptstyle e},\nu_{\scriptscriptstyle \mu}, 
\nu_{\scriptscriptstyle \tau})$.
The symmetric, but generally complex, matrix can be diagonalized 
by using unitary matrix {\boldmath $X$} such that
\beq
\mbox{\boldmath $X$}^{\!\scriptscriptstyle  T} 
{\cal M_{\scriptscriptstyle N}}\mbox{\boldmath $X$} =
\mbox{diag} \{ {M}_{\!\scriptscriptstyle \chi^0_{n}} \} \; .
\eeq
Again, the first part of the mass eigenvalues, 
${M}_{\!\scriptscriptstyle \chi^0_{n}} $ for $n = 1$--$4$ here, gives
the heavy states, {\it i.e.} neutralinos. The last part, 
${M}_{\!\scriptscriptstyle \chi^0_{n}} $ for $n = 5$--$7$ are hence
physical neutrino masses at tree-level. 

Consider the mass matrix in the form of $4+3$ block submatrices:
\begin{equation} \label{mnuA}
{\cal M_N} = \left( \begin{array}{cc}
              {\cal M}_n & \xi^{\!\scriptscriptstyle  T} \\
              \xi & m_\nu^o \end{array}  \right ) \;.
\end{equation}
In the interest of small neutrino masses, a perturbative (seesaw) 
block diagonalization can be applied. Explicitly, the diagonalizing 
matrix can be written approximately as
\[
\mbox{\boldmath $Z$}\simeq
\left( \begin{array}{cc}
I_{\!\scriptscriptstyle 4 \times 4} & ({\cal M}_n^{\mbox{-}1} \,  \xi^{\!\scriptscriptstyle  T}) \\
-({\cal M}_n^{\mbox{-}1} \,  \xi^{\!\scriptscriptstyle  T})^\dag  &  I_{\!\scriptscriptstyle 3 \times 3}
\end{array} \right) \; .
\]
The tree-level effective neutrino mass matrix  can be then obtained as 
\beqa \label{ssA}
(m_\nu)  \simeq 
&& 
- ({\cal M}_n^{\mbox{-}1} \,  \xi^{\!\scriptscriptstyle  T})
^{\scriptscriptstyle T} \, {\cal M}_n \, ({\cal M}_n^{\mbox{-}1} \,  \xi^{\!\scriptscriptstyle  T})
= - \, \xi \, {\cal M}_n^{-1} \, \xi^{\!\scriptscriptstyle T} \; 
\nonumber \\
\simeq &&   \frac{  M_{\!\scriptscriptstyle Z}^2 \, \cos^2\!\!\beta \,
 (M_{\!\scriptscriptstyle 1} \, \cos^2\!\theta_{\!\scriptscriptstyle W} 
+ M_{\!\scriptscriptstyle 2} \, \sin^2\!\theta_{\!\scriptscriptstyle W} ) \;} 
 { \det({\cal M}_n) }  \;\;
(\, \mu_i \, \mu_j \,) \;\;,
\eeqa
where
\begin{eqnarray}
({\cal M}_n^{\mbox{-}1} \,  \xi^{\!\scriptscriptstyle  T})_{1j}
 &=& -\mu_i \; \frac{M_{\!\scriptscriptstyle Z} \cos\!\zb \;
\mu_{\scriptscriptstyle 0} M_{\!\scriptscriptstyle 2} 
\sin\! \theta_{\!\scriptscriptstyle W}}{\det({\cal M}_n)} 
\;, \nonumber \\
 ({\cal M}_n^{\mbox{-}1} \,  \xi^{\!\scriptscriptstyle  T})_{2j}
 &=& \mu_i \; \frac{M_{\!\scriptscriptstyle Z} \cos\!\zb \;
\mu_{\scriptscriptstyle 0}  M_{\!\scriptscriptstyle 1} 
\cos\! \theta_{\!\scriptscriptstyle W}}{\det({\cal M}_n)} 
\;, \nonumber \\
({\cal M}_n^{\mbox{-}1} \,  \xi^{\!\scriptscriptstyle  T})_{3j}
 &=& \mu_i \; \frac{M_{\!\scriptscriptstyle Z}^2  \cos^2\!\!\zb \; 
(M_{\!\scriptscriptstyle 1} \cos\!^2 \theta_{\!\scriptscriptstyle W}
 + M_{\!\scriptscriptstyle 2} \sin\!^2 \theta_{\!\scriptscriptstyle W})}
{\det({\cal M}_n)} 
\;, \nonumber \\
 ({\cal M}_n^{\mbox{-}1} \,  \xi^{\!\scriptscriptstyle  T})_{4j}
 &=& - \mu_i \; \frac{\mu_{\scriptscriptstyle 0} \, 
M_{\!\scriptscriptstyle 1} \,  M_{\!\scriptscriptstyle 2} \, 
- M_{\!\scriptscriptstyle Z}^2 \, \sin\!\beta \cos\!\beta \;
 (M_{\!\scriptscriptstyle 1} \cos\!^2 \theta_{\!\scriptscriptstyle W}
 + M_{\!\scriptscriptstyle 2} \sin\!^2 \theta_{\!\scriptscriptstyle W})}
{\det({\cal M}_n)}          \;,
\end{eqnarray}
and
\begin{equation} 
\det({\cal M}_n) = \mu_{\scriptscriptstyle 0} \,
\left[-\mu_{\scriptscriptstyle 0} \,
M_{\!\scriptscriptstyle 1}\, M_{\!\scriptscriptstyle 2} \,
+ M_{\!\scriptscriptstyle Z}^2 \, \sin\!2\beta \,
 ( M_{\!\scriptscriptstyle 1} \, \cos^2\!\theta_{\!\scriptscriptstyle W} 
+ M_{\!\scriptscriptstyle 2} \, \sin^2\!\theta_{\!\scriptscriptstyle W} )
\right] 
\end{equation}
is equivalent in expression to the determinant of the MSSM neutralino mass matrix. 

It is obvious that the $3 \times 3$ matrix $(\,\mu_i \, \mu_j\,)$ has only 
one nonzero eigenvalue given by
\beq
\mu_{\scriptscriptstyle 5}^2 = |\mu_{\scriptscriptstyle 1}|^2 + 
 |\mu_{\scriptscriptstyle 2}|^2 + |\mu_{\scriptscriptstyle 3}|^2 \; .
\eeq
We can define
\beq
R_{\scriptscriptstyle 5} =  \left( \begin{array}{ccc}
\frac{\mu_{\scriptscriptstyle 1}^*}{\mu_{\scriptscriptstyle 5}}
	& 0 & \frac{\sqrt{|\mu_{\scriptscriptstyle 2}|^2 + 
|\mu_{\scriptscriptstyle 3}|^2}}{\mu_{\scriptscriptstyle 5}}
 \\
\frac{\mu_{\scriptscriptstyle 2}^*}{\mu_{\scriptscriptstyle 5}}
 	& \frac{\mu_{\scriptscriptstyle 3}}{\sqrt{|\mu_{\scriptscriptstyle 2}|^2 + |\mu_{\scriptscriptstyle 3}|^2}}
	& - \frac{\mu_{\scriptscriptstyle 1} \, \mu_{\scriptscriptstyle 2}^*}
{\mu_{\scriptscriptstyle 5} \, \sqrt{|\mu_{\scriptscriptstyle 2}|^2 + |\mu_{\scriptscriptstyle 3}|^2}}
 \\
\frac{\mu_{\scriptscriptstyle 3}^*}{\mu_{\scriptscriptstyle 5}}
	& -\frac{\mu_{\scriptscriptstyle 2}}{\sqrt{|\mu_{\scriptscriptstyle 2}|^2 + |\mu_{\scriptscriptstyle 3}|^2}}
	& - \frac{\mu_{\scriptscriptstyle 1} \, \mu_{\scriptscriptstyle 3}^*}
{\mu_{\scriptscriptstyle 5} \, \sqrt{|\mu_{\scriptscriptstyle 2}|^2 + |\mu_{\scriptscriptstyle 3}|^2}}
\end{array} \right) \; .
\eeq
Then, we have  $R_{\scriptscriptstyle 5}^{\scriptscriptstyle  T} \, 
(\,\mu_i \, \mu_j\,) \, R_{\scriptscriptstyle 5} = \rm{diag}\{\,
\mu_{\scriptscriptstyle 5}^2, 0, 0 \, \}$. Here, $\mu_{\scriptscriptstyle 5}$ and
$\sqrt{|\mu_{\scriptscriptstyle 2}|^2 + |\mu_{\scriptscriptstyle 3}|^2}$ are taken
as real and positive. With this result, we can write the overall diagonalizing 
matrix {\boldmath $X$} in the form
\beq
\mbox{\boldmath $X$} \simeq
\left( \begin{array}{cc}
I_{\!\scriptscriptstyle 4 \times 4} &  ({\cal M}_n^{\mbox{-}1} \, 
\xi^{\!\scriptscriptstyle  T}) \\
- ({\cal M}_n^{\mbox{-}1} \, 
\xi^{\!\scriptscriptstyle  T})^\dag  &  I_{\!\scriptscriptstyle 3 \times 3}
\end{array} \right) \,
\left( \begin{array}{cc}
R_{\scriptscriptstyle n} &  \quad  0_{\scriptscriptstyle 4 \times 3} \\
0_{\scriptscriptstyle 3 \times 4}  &    \quad e^{i\zeta} \, R_{\scriptscriptstyle 5}
\end{array} \right)
= \left( \begin{array}{cc}
R_{\scriptscriptstyle n} &   \quad e^{i\zeta} \,  ({\cal M}_n^{\mbox{-}1} \, 
\xi^{\!\scriptscriptstyle  T}) \,  R_{\scriptscriptstyle 5}\\
- ({\cal M}_n^{\mbox{-}1} \, 
\xi^{\!\scriptscriptstyle  T})^\dag \, R_{\scriptscriptstyle n} &     \quad e^{i\zeta} \, R_{\scriptscriptstyle 5}
\end{array} \right) \; ,
\eeq
where $R_{\scriptscriptstyle n}$ is a $4 \times 4$ matrix with elements all
expected to be of order 1, basically the diagonalizing matrix for the ${\cal M}_n$
block and $e^{i\zeta}$ is a constant phase factor put in to absorb the overall
phase in the constant factor in the expression of Eq.(\ref{ssA}) so that the
resulted neutrino mass eigenvalue would be real and positive. The matrix 
{\boldmath $X$} contains the important information of the gaugino and higgsino 
contents of the physical neutrinos. This is given by the mixing elements in the 
off-diagonal blocks. The  {\boldmath $Z$} matrix in itself gives similar information
for the effective SM neutrinos (flavor states). The latter matrix may be more useful
in the analysis of neutrino phenomenology.

\section{The scalar sectors} 
The SVP also simplifies much the otherwise extremely complicated expressions
for the (tree-level) mass-squared matrices of the scalar sectors. Scalars with the 
same color and electric charges may, of course, mix with one another. The story
gets a bit complicated with, again, the color-singlet scalars. Here, there are
nine physical neutral scalars and seven physical charged scalars, together with
a unphysical Goldstone state in each case. In the MSSM, these are separated
into the Higgses and the sleptons. The separation is no longer valid in the GSSM.
Recall that under the SVP we can still identify the Higgses as states that come 
from the superfield multiplets $\hat{H}_u$ and $\hat{L}_0$. The physical states,
however, are expected to be mixture of the Higgses and the sleptons. Similar to
the case for the fermions discussed in the previous section, we will be particularly
interested in mixings between the Higgses and the other sleptons. We will also give
explicit perturbative formulae for such mixings in the small neutrino mass scenario.
Another thing we will address is the $LR$ squark and slepton mixings, which also
have interesting contributions beyond those in the MSSM. We start first with 
the squarks sectors. Note that in all the expressions given in this section, we
neglect contributions from the nonholomorphic soft terms [{\it cf.} Eq.(\ref{nhsoft})]. 
Such contributions go only into the $LR$-mixing part, in exactly the same way as they 
do in the MSSM. Their explicit incorporation is hence straightforward. 

\subsection{The squarks}
The up-squark mass-squared matrix looks exactly as the one in the MSSM, hence we skip
here. The down-squark sector, however, has interesting result. 
We have the mass-squared matrix as follows : 
\beq \label{MD}
{\cal M}_{\!\scriptscriptstyle {D}}^2 =
\left( \begin{array}{cc}
{\cal M}_{\!\scriptscriptstyle LL}^2 & {\cal M}_{\!\scriptscriptstyle RL}^{2\dag} \\
{\cal M}_{\!\scriptscriptstyle RL}^{2} & {\cal M}_{\!\scriptscriptstyle RR}^2
 \end{array} \right) \; ,
\eeq
where
\beqa
{\cal M}_{\!\scriptscriptstyle LL}^2 =
\tilde{m}_{\!\scriptscriptstyle {Q}}^2 +
m_{\!\scriptscriptstyle D}^\dag m_{\!\scriptscriptstyle D}
+ M_{\!\scriptscriptstyle Z}^2\, \cos\!2 \beta 
\left[ -\frac{1}{2} + \frac{1}{3} \sin\!^2 \theta_{\!\scriptscriptstyle W}\right] 
I_{\scriptscriptstyle 3 \times 3} \; ,
\nonumber \\
{\cal M}_{\!\scriptscriptstyle RR}^2 =
\tilde{m}_{\!\scriptscriptstyle {D}}^2 +
m_{\!\scriptscriptstyle D} m_{\!\scriptscriptstyle D}^\dag
+ M_{\!\scriptscriptstyle Z}^2\, \cos\!2\beta 
\left[  - \frac{1}{3} \sin\!^2 \theta_{\!\scriptscriptstyle W}\right] 
I_{\scriptscriptstyle 3 \times 3} \; ,
\eeqa
and
\beqa 
({\cal M}_{\!\scriptscriptstyle RL}^{2})^{\scriptscriptstyle T} 
&=& 
A^{\!{\scriptscriptstyle D}} \frac{v_{\scriptscriptstyle 0}}{\sqrt{2}}
- (\, \mu_{\scriptscriptstyle \za}^*\, \lambda^{\!\prime}_{{\scriptscriptstyle \za}jk}\,)
\; \frac{v_{\scriptscriptstyle u}}{\sqrt{2}} \; 
\nonumber \\ &=& 
\left[ A_d -  \mu_{\scriptscriptstyle 0}^* \, \tan\!\beta \right]
\,m_{\!\scriptscriptstyle D}\;
+ \frac{\sqrt{2}\, M_{\!\scriptscriptstyle W} \cos\!\beta}
{g_{\scriptscriptstyle 2} } \,
\delta\! A^{\!{\scriptscriptstyle D}}
- \frac{\sqrt{2}\, M_{\!\scriptscriptstyle W} \sin\!\beta}
{g_{\scriptscriptstyle 2} } \,
(\, \mu_i^* \, \lambda^{\!\prime}_{ijk}\, ) \; .
\label{RL}
\eeqa
Here, $m_{\!\scriptscriptstyle D}$ is the down-quark mass matrix, 
which is diagonal under the parametrization adopted; 
$A_d$ is a constant (mass) parameter representing the 
``proportional" part of the $A$-term and the matrix 
$\delta\! A^{\!{\scriptscriptstyle D}}$ is the ``proportionality" violating 
part; $(\, \mu_i^* \, \lambda^{\!\prime}_{ijk}\, )$, and similarly 
$(\, \mu_{\scriptscriptstyle \za}^* \, \lambda^{\!\prime}_{{\scriptscriptstyle \za}jk}\,)$, 
denotes the $3\times 3$ matrix $(\;)_{jk}$ with elements listed. 
$\!\!$\footnote{Note that we use this kind of bracket notations for matrices 
extensively here. In this case, the repeated index $i$ is to be summed 
over as usual, and hence dummy.}
The  $(\, \mu_{\scriptscriptstyle \za}^* \, \lambda^{\!\prime}_{{\scriptscriptstyle \za}jk}\,)$
term is the full $F$-term contribution, while the 
$(\, \mu_i^* \, \lambda^{\!\prime}_{ijk}\, )$ part separated out in the last 
expression gives the new contributions beyond that of the MSSM.  The term actually
gives new contributions to the quark electric dipole moment, for example, as
discussed in Refs\cite{as4,as6}.

\subsection{The neutral scalars}
We have five neutral complex scalar fields, all from electroweak doublets.
They are the $\hat{H}_u$ and the four $\hat{L}_\za$'s. Explicitly, we write the 
$(1+4)$ complex field,  $\phi_n$'s, in the order
$(\,{h}_{\!\scriptscriptstyle u}^{{\!\scriptscriptstyle 0}\dag}\,, 
\tilde{l}_{\scriptscriptstyle 0}^{\scriptscriptstyle 0}\,, 
\tilde{l}_{\scriptscriptstyle 1}^{\scriptscriptstyle 0}\,,
\tilde{l}_{\scriptscriptstyle 2}^{\scriptscriptstyle 0}\,,
\tilde{l}_{\scriptscriptstyle 3}^{\scriptscriptstyle 0}\,)$.
Using this basis, all the neutral scalar mass terms can be written in two parts
--- a simple $({\cal M}_{\!\scriptscriptstyle {\phi}{\phi}^{\!\dag}}^2)_{mn} \,
\phi_m^\dag \phi_n$ part, and a Majorana-like part in the form 
${1\over 2} \,  ({\cal M}_{\!\scriptscriptstyle {\phi\phi}}^2)_{mn} \,
\phi_m \phi_n + \mbox{h.c.}$. As the neutral scalars are originated
from chiral doublet superfields, the existence of the Majorana-like
part is a direct consequence of the electroweak symmetry
breaking VEVs, hence restricted to the scalars playing the Higgs
role only. They come from the quartic terms of the Higgs fields in
the scalar potential. We have, explicitly,
\begin{eqnarray}
 \label{Mpp}
{\cal M}_{\!\scriptscriptstyle {\phi\phi}}^2 =
{1\over 2} \, M_{\!\scriptscriptstyle Z}^2\,
\left( \begin{array}{ccc}
 \sin\!^2\! \beta  &  - \cos\!\beta \, \sin\! \beta
& \quad 0_{\scriptscriptstyle 1 \times 3} \\
 - \cos\!\beta \, \sin\! \beta & \cos\!^2\! \beta 
& \quad 0_{\scriptscriptstyle 1 \times 3} \\
0_{\scriptscriptstyle 3 \times 1} & 0_{\scriptscriptstyle 3 \times 1} 
& \quad 0_{\scriptscriptstyle 3 \times 3} 
\end{array} \right) \; ;
\end{eqnarray}
and
\begin{eqnarray}
{\cal M}_{\!\scriptscriptstyle {\phi}{\phi}^{\!\dag}}^2 &=&
{\cal M}_{\!\scriptscriptstyle {\phi}}^2 
+  {\cal M}_{\!\scriptscriptstyle {\phi\phi}}^2 \; ,
\label{Mp}
\end{eqnarray}
where
\beq \label{Mps}
{\cal M}_{\!\scriptscriptstyle {\phi}}^2 =
\left( \begin{array}{cc}
\tilde{m}_{\!\scriptscriptstyle H_{\!\scriptscriptstyle u}}^2
+ \mu_{\!\scriptscriptstyle \za}^* \mu_{\scriptscriptstyle \za}
+ M_{\!\scriptscriptstyle Z}^2\, \cos\!2 \beta 
\left[-\frac{1}{2}\right] 
& - (B_\za) \\
- (B_\za^*) &
\tilde{m}_{\!\scriptscriptstyle {L}}^2 
+ (\mu_{\!\scriptscriptstyle \za}^* \mu_{\scriptscriptstyle \zb})
+ M_{\!\scriptscriptstyle Z}^2\, \cos\!2 \beta 
\left[ \frac{1}{2}\right]   I_{\scriptscriptstyle 4 \times 4} 
\end{array} \right) \; .
\eeq
Note that ${\cal M}_{\!\scriptscriptstyle {\phi\phi}}^2$ here is 
real, due to results from Sec.~II~D above; while 
${\cal M}_{\!\scriptscriptstyle {\phi}{\phi}^{\!\dag}}^2$ does have complex entries.
Writing the five $\phi_n$'s in terms of their scalar and pseudoscalar parts,
the full $10\times 10$ (real and symmetric) mass-squared matrix for 
the real scalars is then given by
\begin{equation} \label{MSN}
{\cal M}_{\!\scriptscriptstyle S}^2 =
\left( \begin{array}{cc}
{\cal M}_{\!\scriptscriptstyle SS}^2 &
{\cal M}_{\!\scriptscriptstyle SP}^2 \\
({\cal M}_{\!\scriptscriptstyle SP}^{2})^{\!\scriptscriptstyle T} &
{\cal M}_{\!\scriptscriptstyle PP}^2
\end{array} \right) \; ,
\end{equation}
where the scalar, pseudoscalar, and mixing parts are
\begin{eqnarray}
{\cal M}_{\!\scriptscriptstyle SS}^2 &=&
\mbox{Re}({\cal M}_{\!\scriptscriptstyle {\phi}{\phi}^{\!\dag}}^2)
+ {\cal M}_{\!\scriptscriptstyle {\phi\phi}}^2 
= \mbox{Re}({\cal M}_{\!\scriptscriptstyle {\phi}}^2)
+ 2\, {\cal M}_{\!\scriptscriptstyle {\phi\phi}}^2 \; ,
\nonumber \\
{\cal M}_{\!\scriptscriptstyle PP}^2 &=&
\mbox{Re}({\cal M}_{\!\scriptscriptstyle {\phi}{\phi}^{\!\dag}}^2)
- {\cal M}_{\!\scriptscriptstyle {\phi\phi}}^2 
= \mbox{Re}({\cal M}_{\!\scriptscriptstyle {\phi}}^2) \; ,
\nonumber \\
{\cal M}_{\!\scriptscriptstyle SP}^2 &=& - 
 \mbox{Im}({\cal M}_{\!\scriptscriptstyle {\phi}{\phi}^{\!\dag}}^2) 
= - \mbox{Im}({\cal M}_{\!\scriptscriptstyle {\phi}}^2) \; ,
\label{lastsc} 
\end{eqnarray} 
respectively. If $\mbox{Im}({\cal M}_{\!\scriptscriptstyle {\phi}}^2)$
vanishes, the scalars and pseudoscalars decouple from one another and 
the unphysical Goldstone mode would be found among the latter. 
Note that our expansion of the $\phi_n$'s into scalar 
and pseudoscalar parts here takes the universal form 
$\phi_n = {1\over \sqrt{2}}(\phi_{n}^s + i \phi_{n}^a)$, hence we actually have
${h}_{\!\scriptscriptstyle u}^{\!\scriptscriptstyle 0}
=  {1\over \sqrt{2}}({h}_{\!\scriptscriptstyle u}^s 
-i {h}_{\!\scriptscriptstyle u}^a)$.

As a real scalar mass matrix, ${\cal M}_{\!\scriptscriptstyle S}^2$
could be diagonalized by an orthogonal matrix ${\cal D}^{s}$. However, it is
sometimes useful to consider ${\cal D}^{s}$ as if it is just an unitary matrix.  
Thinking about the neutral scalars as complex scalars instead of in terms of the 
scalar and pseudoscalar constituents also helps to illustrate some theoretical
features. These considerations are especially valid for the three
$\tilde{l}_i^{\scriptscriptstyle 0}$'s, which are usually called ``sneutrino".
$\!\!\!$\footnote{
They are not exactly the scalar partners of the physical neutrinos.
}
Hence, we write ${\cal D}^{s\dag}  {\cal M}_{\!\scriptscriptstyle S}^2 \, 
{\cal D}^{s} = \mbox{diag}\{\,M_{\!\scriptscriptstyle S_m}^2, m=1\,\mbox{to}\,10\,\}$.
It is useful to consider the form of ${\cal D}^{s}$ closest to the identity
matrix, {\it i.e.,} with all diagonal entries being order one. The unphysical 
Goldstone mode has, of course, to be found then among the first two
pseudoscalars. The mode is naturally label as the $m=6$ mass eigenstate here.
All the off-diagonal entries except those related 
to mixing of the Higgses ({\it i.e.} the 12-, 21-, 67-, and 76-entries)
are expected to be relatively small.
 
Now we want to decouple the unphysical pseudoscalar explicitly. Note that
${\cal M}_{\!\scriptscriptstyle {\phi}}^2$
can be rewritten, through using the tadpole equations (\ref{tp}) and (\ref{tp3})
as
\beq  \label{MphiB}
{\cal M}_{\!\scriptscriptstyle {\phi}}^2 =
\left( \begin{array}{ccc}
B_0 \, \cot\!\zb & -B_0 & -(B_i) \\
-B_0 & B_0 \, \tan\!\zb & (B_i) \, \tan\!\zb\\
-(B_i^*) & (B_i^*) \, \tan\!\zb & (\star)
\end{array} \right)\; ,
\eeq
with $B_0$ taken as real; and the $(\star)$ denotes the last $3\times 3$ block  
in the original form. The matrix can be diagonalized by a simple rotation among
the first two states given by
\beq \label{Rb}
R_\zb = \left( \begin{array}{cc}
\sin\!\zb &  -\cos\!\zb   \\
\cos\!\zb & \sin\!\zb    
\end{array} \right) \; .
\eeq
Explicitly, we have
\beq
\left( \begin{array}{cc}
R_\zb^{\!\scriptscriptstyle T} &  0_{\scriptscriptstyle 2 \times 3}   \\
0_{\scriptscriptstyle 3 \times 2}  &   I_{\!\scriptscriptstyle 3 \times 3} 
\end{array} \right) \;
{\cal M}_{\!\scriptscriptstyle {\phi}}^2 \; 
\left( \begin{array}{cc}
R_\zb &  0_{\scriptscriptstyle 2 \times 3}   \\
0_{\scriptscriptstyle 3 \times 2}  &   I_{\!\scriptscriptstyle 3 \times 3} 
\end{array} \right) = 
\left( \begin{array}{ccc}
\; 0 \; & \quad 0 \quad & 0 \\
0 & \frac{2\,B_0}{\sin\!2\zb} & \frac{1}{\cos\!\zb} (B_i)\\
0 &  \frac{1}{\cos\!\zb} (B_i^*)& (\star)
\end{array} \right) \; ,
\eeq
where the $3\times 3$ block denotes by $(\star)$ is left untouched. The extended
rotation given by $\mbox{diag}\{\,I_{\!\scriptscriptstyle 5 \times 5} , 
R_\zb, I_{\!\scriptscriptstyle 3 \times 3} \,\}$ then obviously decouples the 
resulted $6$-th state as the massless unphysical mode.

Next, we introduce 
\beq \label{Ra}
R_\za = \left( \begin{array}{cc}
\cos\!\za & -\sin\!\za    \\
\sin\!\za &  \cos\!\za 
\end{array} \right) \; ,
\eeq
as the diagonalizing matrix for the first $2\times 2$ block of 
${\cal M}_{\!\scriptscriptstyle SS}^2$. Define 
$R_{\za\zb}= \mbox{diag} \{ \, R_\za,\, I_{\scriptscriptstyle 3 \times 3},\,
R_\zb,\, I_{\scriptscriptstyle 3 \times 3}\,\}$.  Then, the matrix 
$R_{\za\zb}^{\!\scriptscriptstyle T}\,{\cal M}_{\!\scriptscriptstyle S}^2\,R_{\za\zb}$ 
has only small off-diagonal entries arising from two
sources. One of the latter is the set of lepton-flavor mixing soft masses familiar
in the MSSM, namely, the $\tilde{m}_{\!\scriptscriptstyle {L}_{ij}}^2$'s
in our notation; the other is the set of terms dependent on the $\mu_i$'s, the 
$B_i$'s, and the $\tilde{m}_{\!\scriptscriptstyle {L}_{0i}}^2$'s. 
In fact, using Eq.(\ref{MphiB}), we can see that 
$R_{\za\zb}^{\!\scriptscriptstyle T} \, {\cal M}_{\!\scriptscriptstyle S}^2 \, R_{\za\zb}$
can be written as
\footnotesize \[
\left( \begin{array}{cccccc}
M_{\!\scriptscriptstyle S_{1'}}^2 & 0 
& (\mbox{Re}[B_i]) \, [\tan\!\zb \, \sin\!\za - \cos\!\za]
& 0 & 0 & -(\mbox{Im}[B_i]) \, [\tan\!\zb \, \sin\!\za - \cos\!\za]
\\
 & M_{\!\scriptscriptstyle S_{2'}}^2 
& (\mbox{Re}[B_i]) \, [\tan\!\zb \, \cos\!\za + \sin\!\za]
& 0 & 0 & -(\mbox{Im}[B_i]) \, [\tan\!\zb \, \cos\!\za + \sin\!\za]
\\
 &  & (\mbox{Re}[\tilde{m}_{\!\scriptscriptstyle {L}_{ij}}^2 + \mu_i^*\mu_j])
+ {1\over 2}\, M_{\!\scriptscriptstyle Z}^2  \cos\!2\zb \, I_{\scriptscriptstyle 3 \times 3}
& 0 & \frac{1}{\cos\!\zb} (\mbox{Im}[B_i])
&  -(\mbox{Im}[\tilde{m}_{\!\scriptscriptstyle {L}_{ij}}^2 + \mu_i^*\mu_j])
\\
 &  & & 0 & 0 & 0 \\
 &  & &   & \frac{2\,B_0}{\sin\!2\zb} 
& \frac{1}{\cos\!\zb} (\mbox{Re}[B_i])\\
 &  &  &  &    
& (\mbox{Re}[\tilde{m}_{\!\scriptscriptstyle {L}_{ij}}^2 + \mu_i^*\mu_j])
+ {1\over 2}\, M_{\!\scriptscriptstyle Z}^2  \cos\!2\zb  \, I_{\scriptscriptstyle 3 \times 3}
\end{array} \right) \; ,
\] \normalsize
where we have explicitly written out only the elements in the upper triangular part
(of the symmetric matrix), and introduced 
$M_{\!\scriptscriptstyle S_{1'}}^2$ and $M_{\!\scriptscriptstyle S_{2'}}^2$ 
to denote the 11- and 22- elements. The latter notation is based on the fact that
the two diagonal entries would be approximately the mass eigenvalues
$M_{\!\scriptscriptstyle S_{1}}^2$ and $M_{\!\scriptscriptstyle S_{2}}^2$,
respectively, for small $B_i$'s. Again, the 6-th state is the decoupled 
unphysical Goldstone mode.

We are now ready to introduce the diagonalizing matrix ${\cal D}^{s}$ to the
neutral scalars
${\cal D}^{s^{\!\scriptscriptstyle T}} \!\! {\cal M}_{\!\scriptscriptstyle S}^2 \, {\cal D}^{s}
= \mbox{diag}\{\,M_{\!\scriptscriptstyle S_m}^2, m=1\,\mbox{to}\,10\,\}$
written as 
\[
{\cal D}^{s} \equiv R_{\za\zb} \, R^s \;.
\]
In the phenomenologically interesting case with all the 
off-diagonal entries to the matrix 
$R_{\za\zb}^{\!\scriptscriptstyle T}\,{\cal M}_{\!\scriptscriptstyle S}^2\,R_{\za\zb}$
being  small, one can easily obtain useful expressions for the interesting 
diagonalizing matrix elements. In particular,  we have 
\beqa
{\cal D}^{s}_{\!(i+2)m} =R^{s}_{(i+2)m} \; ,
\nonumber \\
{\cal D}^{s}_{\!(i+7)m} =R^{s}_{(i+7)m} \; ,
\eeqa
which can be read out from implementing a perturbative diagonalizing formula
on the mass matrix
$R_{\za\zb}^{\!\scriptscriptstyle T} \, {\cal M}_{\!\scriptscriptstyle S}^2 \, R_{\za\zb}$
as explicitly given above. Furthermore, we have
\beqa
{\cal D}^{s}_{\!1(i+2)} & \simeq &
\frac{-\mbox{Re}[B_i]}{M_{\!s}^2} \; ,
\nonumber \\
{\cal D}^{s}_{\!1(i+7)} & \simeq &
\frac{\mbox{Im}[B_i]}{M_{\!s}^2} \; ,
\eeqa
and
\beqa
{\cal D}^{s}_{\!2(i+2)} & \simeq &
\frac{\mbox{Re}[B_i] \, \tan\!\zb}{M_{\!s}^2} \; ,
\nonumber \\
{\cal D}^{s}_{\!2(i+7)} & \simeq &
\frac{-\mbox{Im}[B_i] \, \tan\!\zb}{M_{\!s}^2} \; ,
\eeqa
where we introduce $M_{\!s}^2$ to denote a generic mass-squared
parameter at the scalar mass-squared (or $\tilde{m}_{\!\scriptscriptstyle {L}}^2$)
scale; in particular, here it represents  the quantities
$\Big[ \tilde{m}_{\!\scriptscriptstyle {L}_{ii}}^2 +|\mu_i|^2
+ {1\over 2}\,M_{\!\scriptscriptstyle Z}^2\, \cos\!2 \beta 
-M_{\!\scriptscriptstyle S_{1'}}^2\Big]$ and 
$\Big[ \tilde{m}_{\!\scriptscriptstyle {L}_{ii}}^2 +|\mu_i|^2
+ {1\over 2}\,M_{\!\scriptscriptstyle Z}^2\, \cos\!2 \beta 
-M_{\!\scriptscriptstyle S_{2'}}^2\Big]$ respectively.
Similarly, we have
\beqa
{\cal D}^{s}_{\!6(i+2)} & \simeq &
\frac{-\mbox{Im}[B_i]}{M_{\!s}^2} \; ,
\nonumber \\
{\cal D}^{s}_{\!6(i+7)} & \simeq &
\frac{-\mbox{Re}[B_i]}{M_{\!s}^2} \; ,
\nonumber \\
{\cal D}^{s}_{\!7(i+2)} & \simeq &
\frac{\mbox{Im}[B_i] \, \tan\!\zb}{M_{\!s}^2} \; ,
\nonumber \\
{\cal D}^{s}_{\!7(i+7)} & \simeq &
\frac{\mbox{Re}[B_i] \, \tan\!\zb}{M_{\!s}^2} \; ,
\eeqa
with $M_{\!s}^2$ representing
$\Big[ \tilde{m}_{\!\scriptscriptstyle {L}_{ii}}^2 +|\mu_i|^2
+ {1\over 2}\,M_{\!\scriptscriptstyle Z}^2\, \cos\!2 \beta 
- \frac{2\,B_0}{\sin\!2\zb} \Big]$ and
$\Big[ \tilde{m}_{\!\scriptscriptstyle {L}_{ii}}^2 +|\mu_i|^2
+ {1\over 2}\,M_{\!\scriptscriptstyle Z}^2\, \cos\!2 \beta 
- \frac{2\,B_0}{\sin\!2\zb} \Big]$ respectively. Note that one would like to write
the ${\cal D}^{s}_{\!(i+2)m}$ and ${\cal D}^{s}_{\!(i+7)m}$ matrix elements in a
similar form, {\it e.g.} we write
\beq
{\cal D}^{s}_{\!(i+2)1} \simeq
\frac{-\mbox{Re}[B_i]}{M_{\!s}^2}  \, (\tan\!\zb \, \sin\!\za - \cos\!\za) \; .
\eeq

\subsection{The charged scalars}
From Eq.(\ref{soft}) above, we 
can see that the charged Higgses should be
considered on the same footing together with the sleptons. We have hence an
$8\times 8$ mass-squared matrix. We use the 
basis 	$\{\, h_u^{{\!\scriptscriptstyle +}\dag}, 
\tilde{l}_{\scriptscriptstyle 0}^{\!\!\mbox{ -}},
\tilde{l}_{\scriptscriptstyle 1}^{\!\!\mbox{ -}},
\tilde{l}_{\scriptscriptstyle 2}^{\!\!\mbox{ -}},
\tilde{l}_{\scriptscriptstyle 3}^{\!\!\mbox{ -}},
\tilde{l}_{\scriptscriptstyle 1}^{{\!\scriptscriptstyle +}\dag},
\tilde{l}_{\scriptscriptstyle 2}^{{\!\scriptscriptstyle +}\dag},
\tilde{l}_{\scriptscriptstyle 3}^{{\!\scriptscriptstyle +}\dag}
\,\}$
to write the mass-squared matrix in the following $1+4+3$ form :
\beq \label{ME}
{\cal M}_{\!\scriptscriptstyle {E}}^2 =
\left( \begin{array}{ccc}
\widetilde{\cal M}_{\!\scriptscriptstyle H\!u}^2 &
\widetilde{\cal M}_{\!\scriptscriptstyle LH}^{2\dag}  & 
\widetilde{\cal M}_{\!\scriptscriptstyle RH}^{2\dag}
\\
\widetilde{\cal M}_{\!\scriptscriptstyle LH}^2 & 
\widetilde{\cal M}_{\!\scriptscriptstyle LL}^{2} & 
\widetilde{\cal M}_{\!\scriptscriptstyle RL}^{2\dag} 
\\
\widetilde{\cal M}_{\!\scriptscriptstyle RH}^2 &
\widetilde{\cal M}_{\!\scriptscriptstyle RL}^{2} & 
\widetilde{\cal M}_{\!\scriptscriptstyle RR}^2  
\end{array} \right) \; ;
\eeq
where
\beqa
\widetilde{\cal M}_{\!\scriptscriptstyle H\!u}^2 &=&
\tilde{m}_{\!\scriptscriptstyle H_{\!\scriptscriptstyle u}}^2
+ \mu_{\!\scriptscriptstyle \za}^* \mu_{\scriptscriptstyle \za}
+ M_{\!\scriptscriptstyle Z}^2\, \cos\!2 \beta 
\left[ \,\frac{1}{2} - \sin\!^2\theta_{\!\scriptscriptstyle W}\right]
\nonumber \\
&+&  M_{\!\scriptscriptstyle Z}^2\,  \sin\!^2 \beta \;
[1 - \sin\!^2 \theta_{\!\scriptscriptstyle W}]  \; ,
\nonumber \\
\widetilde{\cal M}_{\!\scriptscriptstyle LL}^2 &=&
\tilde{m}_{\!\scriptscriptstyle {L}}^2 +
m_{\!\scriptscriptstyle L}^\dag m_{\!\scriptscriptstyle L}
+ (\mu_{\!\scriptscriptstyle \za}^* \mu_{\scriptscriptstyle \zb})
+ M_{\!\scriptscriptstyle Z}^2\, \cos\!2 \beta 
\left[ -\frac{1}{2} +  \sin\!^2 \theta_{\!\scriptscriptstyle W}\right] 
I_{\scriptscriptstyle 4 \times 4} \; ,
\nonumber \\
&+& \left( \begin{array}{cc}
 M_{\!\scriptscriptstyle Z}^2\,  \cos\!^2 \beta \;
[1 - \sin\!^2 \theta_{\!\scriptscriptstyle W}] 
& \quad 0_{\scriptscriptstyle 1 \times 3} \quad \\
0_{\scriptscriptstyle 3 \times 1} & 0_{\scriptscriptstyle 3 \times 3}  
\end{array} \right) \; ,
\nonumber \\
\widetilde{\cal M}_{\!\scriptscriptstyle RR}^2 &=&
\tilde{m}_{\!\scriptscriptstyle {E}}^2 +
m_{\!\scriptscriptstyle E} m_{\!\scriptscriptstyle E}^\dag
+ M_{\!\scriptscriptstyle Z}^2\, \cos\!2 \beta 
\left[  - \sin\!^2 \theta_{\!\scriptscriptstyle W}\right] 
I_{\scriptscriptstyle 3 \times 3}
\; ;
\eeqa
and
\beqa 
\widetilde{\cal M}_{\!\scriptscriptstyle LH}^2
&=& (B_{\za}^*)  
+ \left( \begin{array}{c} 
{1 \over 2} \,
M_{\!\scriptscriptstyle Z}^2\,  \sin\!2 \beta \;
[1 - \sin\!^2 \theta_{\!\scriptscriptstyle W}]  \\
0_{\scriptscriptstyle 3 \times 1} 
\end{array} \right)
\; ,
\nonumber \\
\widetilde{\cal M}_{\!\scriptscriptstyle RH}^2
&=&  -\,(\, \mu_i^*\lambda_{i{\scriptscriptstyle 0}k}\, ) \; 
\frac{v_{\scriptscriptstyle 0}}{\sqrt{2}} \; 
= (\, \mu_k^* \, m_k \, ) \hspace*{1in} \mbox{ (no sum over $k$)} \quad \; ,
\nonumber \\
(\widetilde{\cal M}_{\!\scriptscriptstyle RL}^{2})^{\scriptscriptstyle T} 
&=& \left(\begin{array}{c} 
0  \\   A^{\!{\scriptscriptstyle E}} 
\end{array}\right)
 \frac{v_{\scriptscriptstyle 0}}{\sqrt{2}}
 - (\, \mu_{\scriptscriptstyle \za}^*\lambda_{{\scriptscriptstyle \za\zb}k}\, ) \; 
\frac{v_{\scriptscriptstyle u}}{\sqrt{2}} \; 
\nonumber \\
&=& [A_e - \mu_{\scriptscriptstyle 0}^* \, \tan\!\beta ] 
\left(\begin{array}{c}  
0  \\   m_{\!{\scriptscriptstyle E}} 
\end{array}\right) \,
+ \frac{\sqrt{2}\, M_{\!\scriptscriptstyle W} \cos\!\beta}
{g_{\scriptscriptstyle 2} } \,
\left(\begin{array}{c} 
0  \\ \delta\! A^{\!{\scriptscriptstyle E}}
\end{array}\right)
-  \left(\begin{array}{c}  
- \mu_{k}^* \, m_k\, \tan\!\beta \\ 
\frac{\sqrt{2}\, M_{\!\scriptscriptstyle W} \sin\!\beta}
{g_{\scriptscriptstyle 2} } \,(\, \mu_i^*\lambda_{ijk}\, ) 
\end{array}\right) \; .
\label{ERL}
\eeqa
Notations and results here are similar to the squark case above, with some 
difference. We have $A_e$ and $\delta\! A^{\!{\scriptscriptstyle E}}$,
or the extended matrices {\tiny $\left(\begin{array}{c} 
0  \\   \star
\end{array}\right)$} incorporating them, denote the splitting of the $A$-term,
with proportionality defined with respect to 
$m_{\!\scriptscriptstyle E}$; $m_{\!\scriptscriptstyle L}=
\mbox{diag}\{0,m_{\!\scriptscriptstyle E}\}= \mbox{diag}\{0,m_{\!\scriptscriptstyle 1},
m_{\!\scriptscriptstyle 2},m_{\!\scriptscriptstyle 3}\}$. Recall that the $m_i$'s
are approximately the charged lepton masses. 
A $4\times 3$ matrix $(\, \mu_i^*\lambda_{i{\scriptscriptstyle \zb}k}\, )$
gives the new contributions to 
$(\widetilde{\cal M}_{\!\scriptscriptstyle RL}^{2})^{\scriptscriptstyle T}$
beyond that of the MSSM. In the above expression, we separate explicitly the first 
row of the former, which corresponds to mass-squared terms of the type
$\tilde{l}^{\scriptscriptstyle +} h_{\scriptscriptstyle d}^{\!\!\mbox{ -}}$ 
type ($h_{\scriptscriptstyle d}^{\!\!\mbox{ -}} \equiv 
\tilde{l}_{\scriptscriptstyle 0}^{\!\!\mbox{ -}}$). The nonzero 
$\widetilde{\cal M}_{\!\scriptscriptstyle RH}^2$ and the $B_i^*$'s in 
$\widetilde{\cal M}_{\!\scriptscriptstyle LH}^2$ are also interesting new 
contributions. The former is a $\tilde{l}^{\scriptscriptstyle +} 
(h_{\scriptscriptstyle u}^{\scriptscriptstyle +})^{\dag} $
type, while the latter a $\tilde{l}^{\!\!\mbox{ -}} 
h_{\scriptscriptstyle u}^{\scriptscriptstyle +}$  term. Note that the parts with the 
$[1 - \sin\!^2 \theta_{\!\scriptscriptstyle W}]$ factor are singled out 
as they are extra contributions to the masses of the ``charged-Higgses"
({\it i.e.} $\tilde{l}_{\scriptscriptstyle 0}^{\!\!\mbox{ -}}
\equiv h_{\!\scriptscriptstyle d}^{\!\!\mbox{ -}}$ and
$h_{\!\scriptscriptstyle u}^{\!\scriptscriptstyle +}$).
The latter is the result of the quartic terms in the scalar potential and
the fact that the Higgs doublets bear VEVs.

Introducing the diagonalizing matrix ${\cal D}^{l}$, we have
${\cal D}^{l\dag}  {\cal M}_{\!\scriptscriptstyle E}^2 \, {\cal D}^{l}
= \mbox{diag}\{\,M_{\!\scriptscriptstyle \tilde{\ell}_m}^2, m=1\,\mbox{to}\,8\,\}$.
We label the unphysical Goldstone mode by $m=1$. In the small neutrino mass
scenario we are particularly interested in, we expected the 
${\cal M}_{\!\scriptscriptstyle E}^2$ to be dominantly 
diagonal, apart from the mixing between the Higgses ({\it i.e.,}
$h_u$ and $h_d \equiv \tilde{l}_{\scriptscriptstyle 0}$) to give the $m=1$
mode. The matrix ${\cal D}^{l}$ may then be naturally chosen to be close to
identity, {\it i.e.} with all diagonal entries being order 1 and only
the $12$- and $21$-entries being possibly large (order 1) among the 
off-diagonal ones.

The unphysical Goldstone mode is, of course, to be found among the Higgs fields 
${h}_{\!\scriptscriptstyle u}^{\!\scriptscriptstyle +}$ and
${h}_{\!\scriptscriptstyle d}^{\!\!\mbox{ -}} (\equiv
\tilde{l}_{\!\scriptscriptstyle 0}^{\!\!\mbox{ -}})$.
In fact, using the correponding tadpole equations [{\it cf.} Eq.(\ref{tp})],
the first $2\times 2$ (Higgs) block of the matrix
${\cal M}_{\!\scriptscriptstyle {E}}^2$ can be written simply as
\[
\left[\frac{2\,B_0}{\sin\!2\zb} 
+ M_{\!\scriptscriptstyle Z}^2\, \cos\!^2 \theta_{\!\scriptscriptstyle W} \right]
 \left( \begin{array}{cc}
\cos\!^2\zb &  \sin\!\zb \, \cos\!\zb \\
\sin\!\zb \, \cos\!\zb & \sin\!^2\zb
\end{array} \right) \; .
\]
Further using the other tadpole equations [{\it cf.} Eq.(\ref{tp3})], we obtain
\beq
\left( \begin{array}{cc}
R_\zb &  0_{\scriptscriptstyle 2 \times 6}   \\
0_{\scriptscriptstyle 6 \times 2}  &   I_{\!\scriptscriptstyle 6 \times 6} 
\end{array} \right) \,
{\cal M}_{\!\scriptscriptstyle E}^2 \, 
\left( \begin{array}{cc}
R_\zb^{\!\scriptscriptstyle T} &  0_{\scriptscriptstyle 2 \times 6}   \\
0_{\scriptscriptstyle 6 \times 2}  &   I_{\!\scriptscriptstyle 6 \times 6} 
\end{array} \right) \;
= 
\left( \begin{array}{cccc}
0 & 0 & 0 & 0\\
0 & \left[ \frac{2\,B_0}{\sin\!2\zb} 
+ M_{\!\scriptscriptstyle Z}^2\, \cos\!^2 \theta_{\!\scriptscriptstyle W} \right]
& \frac{1}{\cos\!\zb} (B_i) & \frac{1}{\cos\!\zb} (\mu_k \, m_k) \\
0 & \frac{1}{\cos\!\zb} (B_i^*) & \star & \star \\
0 &  \frac{1}{\cos\!\zb} (\mu_k^* \, m_k) & \star & \star
\end{array} \right) \; ,
\eeq
where the $\star$'s denote $3\times 3$ blocks exactly the same as those in
the original ${\cal M}_{\!\scriptscriptstyle E}^2$, details of which
skipped here. Note that there is no sum over $k$  in $(\mu_k \, m_k)$
or  $(\mu_k^* \, m_k)$
 
We are again interested in useful approximate expressions for off-diagonal
elements of the diagonalizing matrix ${\cal D}^{l}$ responsible for mixing between
the Higgses and the other sleptons. From the above, it is easy to obtain
\beqa
{\cal D}^{l}_{1(i+2)}  & \simeq &
\frac{B_i}{M^2_{\!\scriptscriptstyle S}} \; ,
\nonumber \\
{\cal D}^{l}_{2(i+2)}  & \simeq & 
\frac{B_i \, \tan\!\zb}{M^2_{\!\scriptscriptstyle S}} \; ,
\eeqa
where the more exact expression substituted by $M^2_{\!\scriptscriptstyle S}$ here is
$\Big[ \tilde{m}_{\!\scriptscriptstyle {L}_{ii}}^2
+ M_{\!\scriptscriptstyle Z}^2\, \cos\!2 \beta 
\left( -\frac{1}{2} +  \sin\!^2 \theta_{\!\scriptscriptstyle W}\right)
- \left( \frac{2\,B_0}{\sin\!2\zb} 
+ M_{\!\scriptscriptstyle Z}^2\, \cos\!^2 \theta_{\!\scriptscriptstyle W} 
\right)  \Big]$.
Similarly, we have
\beqa
{\cal D}^{l}_{1(i+5)}   & \simeq &
\frac{\mu_i \, m_i}{M^2_{\!\scriptscriptstyle S}} \; ,
\nonumber \\
{\cal D}^{l}_{2(i+5)}   & \simeq &
\frac{\mu_i \, m_i \, \tan\!\zb}{M^2_{\!\scriptscriptstyle S}} \; ,
\eeqa
with $M^2_{\!\scriptscriptstyle S}$ representing
$\Big[ \tilde{m}_{\!\scriptscriptstyle {E}_{ii}}^2
+ M_{\!\scriptscriptstyle Z}^2\, \cos\!2 \beta 
\left( - \sin\!^2 \theta_{\!\scriptscriptstyle W}\right) 
-\left( \frac{2\,B_0}{\sin\!2\zb} 
+ M_{\!\scriptscriptstyle Z}^2\, \cos\!^2 \theta_{\!\scriptscriptstyle W} 
\right) \Big]$.
Note that the elements of the type ${\cal D}^{l}_{\!(i+2)(j+5)}$ are standard 
slepton $LR$ mixing terms with extra contributions [{\it cf.} 
$\widetilde{M}_{\!\scriptscriptstyle {RL}}^2$ from Eq.(\ref{ERL})]. 

\section{Remarks}
We have try to present clearly the formulation of the GSSM, or the (complete)
phenomenological theory of SUSY without R parity in some details above.
We emphasize that the notion of R parity and its violation is a perspective that
looks at the model as an extension of the (R-parity conserving) MSSM.
May be particle physicists are too familiar with the MSSM. The idea of studying
an extension of the MSSM by adding some R-parity violating terms sounds
easy and straightforward. The present approach takes a bit different perspective.
We emphasize studying the GSSM as what it is which we recapitulate again
--- a theory built with the (minimal) superfield spectrum incorporating the SM 
particles, interactions dictated by the SM (gauge) symmetries, and the idea that
SUSY is softly broken. The perspective helps to clarify some confusing issues
within the literature of  R-parity violation.

We advocate strongly the adoption of a specific parametrization, the SVP, which
we elaborate on in this review. The parametrization issue has not been addressed
directly and clearly often enough in the literature. Naively, when the model is
considered as limited versions of extensions of the MSSM, there is no need
to readdress the parametrization beyond the latter framework. Such a perspective,
however, hinders a comprehensive study of the plausible interesting phenomenology.
Our use of the SVP has, for instance, led to the identification of new (RPV) 
tree-level contributions to the superpartner mass matrices and their 
phenomenological implications\cite{as5,as4,as6,as7,as9,as1}. Such features have
been largely overlooked in the literature. The contributions typical involved 
products of bilinear and trilinear (RPV) parameters. The two group of parameters
are simply seldom considered together previously. In our opinion, it is the SVP
framework that renders such features transparent. Moreover, we are only 
beginning to study the detailed phenomenological features of the type. A lot of
work still await our effort.

Among the earlier studies that address the parametrization issue with some care,
the more notable ones are given by Refs.\cite{bh,NP}.  There has been some 
phenomenological use of basically the SVP in Refs.\cite{MP,N}, before we 
advocated explicitly adoption of it as the basic formulation for studying GSSM
(or R-parity violation). Other authors have come to appreciate using the
formulation since then. These include Refs.\cite{GH,AL,DL,CCK}. 

The complete expressions for the mass matrices of fermions and scalars,
together with the perturbative diagonalization expression listed above are
very useful in various phenomenological studies. We again refer interested
readers to our papers on the studies of various phenomenological features
for illustrations\cite{as5,as4,as6,as7,as9}.

\acknowledgements
The author wants to thank M.~Bisset, K.~Cheung, S.-K.~Kang, Y.-Y.~Keum,
C.~Macesanu, L.H.~Orr, for the enjoyable collaborations on the subject
area, and   H.Y.~Cheng, P.H.~Frampton, J.~Feng,
K.~Hagiwara, S.C.~Lee, H.-N.~Li, H.~Murayama, P.~Nath, S.~Pakvasa, 
and X.~Tata for encouragement. Supports from colleagues at University of 
Rochester, and Academia Sinica and National Central University of Taiwan is also 
to be acknowledged. He has also been benefited from activities under the SUSY 
sub-program of the Particle Physics program, National Center for Theoretical 
Sciences, and the hospitality of the center during the period of which part 
of this review is written. His current research is supported by grant 
NSC90-2112-M-008-051 from National Science Council of Taiwan.

\appendix

\section{ Q \& A :-}
In this section, we recapitulate on some aspects of our formulation in
relation to potential confusion from related presentations in the
literature where, in many cases, only partial considerations of R-parity violation
are addressed.\\

\noindent
{\bf \boldmath Why use 4 $\hat{L}_{\alpha}$'s instead of 3 $\hat{L}_i$'s
and one $\hat{H}_{d}$ ?\\
--- because we do not know {\it a priori} to what extend the
Higgs is a superpartner of the charged leptons.} In the MSSM, there is a clearly
enforced distinction between the superfields containing the leptons and the
one containing the Higgs (scalar) doublet responsible for the masses of the
leptons and down-sector quarks. This does not come out naturally from the
minimal superfield spectrum containing the SM. In particular, the distinction
is set by the arbitrarily imposed global symmetry of lepton number.
In fact, we have nowadays, from the neutrino oscillation experiments, strongly 
suggestive evidence on the violation of  lepton number symmetry. Without the
lepton number distinction, we have the four ( $\hat{L}_{\alpha}$) doublet
superfields of the same quantum number which one should not distinguish
{\it a priori}. We have illustrated, in the discussion of the SVP, that there is
a special advantage to identify the Higgs direction among the four doublets,
hence the $\hat{H}_{d}$ notation. One should bear in mind though that the
$\hat{H}_{d}$ ($\equiv   \hat{L}_0$) superfield may contain partly the
charged lepton states. Besides, keeping the four  $\hat{L}_{\alpha}$ notation
helps to keep track of the common ``flavor" structure among the four
doublets. The latter is well illustrated by our discussion of the scalar mass
terms, with pieces that are otherwise easily overlooked.\\

\noindent
{\bf Why should we choose a fixed parametrization (or fixed set of flavor
bases) before doing anything else ?\\
--- because that is the right way to do physics; it gives an unambiguous 
connection between the parameters and the experimental data.} 
Again, we do not do SM physics with quark masses or Yukawa couplings in a 
generic flavor basis. Fixing a parametrization removes redundancy of
parameters. The clearly defined set of parameters then would have a definite
relation to observable physical phenomena. Only then will discussions about
the magnitude of the parameters be sensible. In fact, sets of parameters 
from two different parametrizations, though usually called the same names, 
are not quite the same quantities. They may have different phenomenological 
roles. \\

\noindent
{\bf How about parametrization invariant quantities ? \\
--- only in very limited specific cases could one find quantities of the type that  
may be useful.}
This is particularly the case when there is a high degree of redundancy among
naively defined generic set of parameters with no single parametrization
having any specific advantage. A good example is on the admissible complex
phases of a model Lagrangian. Among all the parameters admitting complex 
phases, in some case, only a small number (combinations of) such phases are
physical. Others could be removed by an optimal parametrization. In the case
of SM quark masses, or the CKM matrix, there is only one physical phase. There,
the Jarkslog invariant used to characterize the resultant CP violating effect, 
gives the only major example of such parametrization invariant quantities in
the literature. Even in that case, the usage is limited. The standard 
parametrization of the CKM matrix (given by the Particle Data Group) does 
use a single particular (arbitrarily chosen) phase. \\

\noindent
{\bf What about parametrization of complex phases under the discussion 
formulation ? \\
--- that still have to be performed.}
Our formulation presented here is admittedly incomplete in this sense. The
issue has basically not been explicitly addressed for any R-parity violation
studies in the literature. However,  one does not expected much redundancy 
among the extra admissible complex phases. The complex phases are of interest 
only in CP violating physics. In most of the other phenomenological studies, only
real parameters are taken. We only started to address interesting new 
contributions to CP violating physics (of fermion electric dipole moments)
recently\cite{as4,as6}. Even in that case, our ignorance of possible redundant 
phases does not hurt much. Nevertheless, the issue of fixing an optimal 
parametrization of the physical phases certainly have to be looked
into carefully when we want to study all the CP violating features of the model in 
good details. \\

\noindent
{\bf \boldmath What other parametrization(s) have been used in the literature ? \\
---    the ``single-$\mu$ parametrization", to some extent.}
The parametrization issue for RPV physics is either not explicitly addressed
or confusingly neglected in many studies in the literature. This is especially
true before our first advocate of the SVP\cite{ru1}. In the case that it
is addressed or a particular parametrization explicitly adopted, it is
usually the single-$\mu$ parametrization introduced almost twenty years
ago\cite{bh}. Interpret under the present notation, the parametrization chooses 
to identify as $\hat{L}_0$, or rather denoted by $\hat{H}_{d}$, the direction
in the space of the  four $\hat{L}_{\alpha}$'s that characterizes the direction
of the $\mu_{\scriptscriptstyle \za}$ couplings. The common way to put it
is that ``the three $\mu_i$'s can be rotated away without loss of generality". 
However, the statement is also a common source of confusion. The first thing
we want to emphasize here is that the  $\mu_i$'s under our formulation
cannot be set to zero. We have an optimal parametrization (apart from a
possible minor redundancy in complex phases) within which no parameter
(generally complex) can be set to vanish without enforcing an extra assumption 
and hence changing the model. All the effects involving the $\mu_i$'s we discussed
in the various papers, with collaborators, are physical. Our formulation,
in our opinion, simply provides the most transparent way to see the 
phenomenological implications. Such physical effects will be described in
terms of different combinations of parameters when a different 
parametrization is used to study the model.

There are a few confusing aspects concerning the explicit or implicit use of the 
single-$\mu$ parametrization that we want to clarify. The parametrization, or
the idea to rotated away the $\mu_i$'s, was first introduced only to study a
limited version of RPV model\cite{bh}. Extending the framework to
include other RPV terms, as many authors did, is a bit less than trivial. When
one rotates away the $\mu_i$'s, one is forced to admit generally nonzero
VEVs for the $\hat{L}_i$'s, often called the sneutrino VEVs. 
Our formulation here, the SVP, chooses to ``rotates away"
the latter, keeping rather nonzero $\mu_i$'s. One certainly cannot do both
at the same time. In the complete model (GSSM) under the single-$\mu$ 
parametrization, the nonzero $\hat{L}_i$ VEVs contribute to masses and
mixings, not only of the neutrinos, but also that of the down-sector quarks
and charged leptons. Together with these VEVs, the $\lambda^{\!\prime}-$ and 
$\lambda-$ type couplings also enter the latter mass matrices.
Then, at least at the conceptually level, one cannot write the Lagrangian with even 
the down-quark superfields, and hence the $\lambda^{\!\prime}-$type
couplings involving them,  in the corresponding mass eigenstate bases.
We have illustrated the misalignment of the charged leptons $\ell_i$ with the  
$\hat{L}_i$ superfields under the SVP. In our case, the story is simple. Each $\mu_i$
characterizes directly such misalignment between an $\ell_i$--$\hat{L}_i$ pair.
It should be obvious that under the single-$\mu$ parametrization,  the situation is
far more complicated, as quite a number of parameters are involved for the
fermion mass terms of within each $\hat{L}_i$. And a similar story goes for 
the down-quark sector. To recapitulate once more,
the SVP allows us the use the down-quark mass eigenstate basis and 
keeps the complication within the leptonic sector, with the whole deviation
from the SM or MSSM setting characterizes only by the three $\mu_i$'s.

Within the use of single-$\mu$ parametrization, there is also the statement
that one can rotate away two of the three ``sneutrino VEVs". This is true.
However, doing that is equivalent to identifying one the $\hat{L}_i$'s
directions, say  $\hat{L}_3$, with the direction of the VEV in the $\hat{L}_i$
space. The catch then is that the chosen $\hat{L}_3$ is generally an arbitrary
linear combination of what may be approximately the $\hat{L}_e$, 
$\hat{L}_\mu$, and $\hat{L}_\tau$. 

Furthermore, to the extent that the single-$\mu$ parametrization has to 
admit nonzero VEVs for the $\hat{L}_i$'s, these superfields have scalar
components that resume some Higgs character though effectively only 
the $\hat{L}_0$ is named a Higgs doublet. The fermion part of the latter
doublet contains partly the physical charged leptons. The physical neutrinos
are the light mass egienstates of a complicated $7\times 7$ neutral  fermion
mass matrix, in any case not just linear combinations of the flavor neutrinos
$\nu_e$, $\nu_\mu$, and $\nu_\tau$. Calling the scalars within the 
$\hat{L}_i$'s sneutrinos is not quite right either.\\

\noindent
{\bf What are the issues involved in going from one parametrization
to another ? \\
--- it is mainly a basis rotation among the superfields; when the VEVs are
involved, it gets a bit complicated. }  For instance, we can certainly take the
SVP as a starting point and perform a $SU(4)$ rotation among the 
$\hat{L}_{\alpha}$'s to a new basis, say, denoted by $\hat{L}_{\alpha}^{\!\prime}$'s 
requiring then the new $\mu_i$'s (couplings of the  
$\hat{H}_u\,\hat{L}_i^{\!\prime}$ terms) be zero. To progress beyond
writing down the Lagrangian in the new basis, one will have to solve for the
scalar potential to find all the nonzero VEVs. The rotation involving only
a basis change among the $\hat{L}_{\alpha}$ superfield now has an effect
also on the interpretation of couplings in the other sectors. The down-quark
superfields are no longer in the mass eigenstate basis (of the tree-level mass 
matrix), as the extra VEVs give off-diagonal contributions discussed above.\\

\noindent
{\bf The bilinear RPV couplings can be rotated away, right?\\
--- depends.} It should be clear from our answers to the above two questions.\\

\noindent
{\bf Should we worry much about doing a basis rotation ?\\
--- generally speaking, we do not have to.}
Physics should be about formulating a theoretical model to be checked
versus experiments. Connecting different formulations of the
same model is not very interesting, unless more than one formulation
have special advantages in specific studies or have been commonly used
(correctly) within the community of physicists. \\

\noindent
{\bf What may be a good parametrization ? \\
--- one that simplifies analyses and enables more direct identification of
the major role of the parameters. } So long as low energy phenomenology is 
concerned, that is all we should care about. We hope that our discussion above
has illustrated the merits of the parametrization presented and advocated
here.  All in all, we urge authors on the subject area to state explicitly the
parametrization adopted and be careful with any extra assumptions used.  
As parameters under the same notation are commonly used under
different parametrization, explicitly stated or otherwise, such assumptions
on some of the parameters have a very different meaning when interpreted
under a different parametrization. This is quite a source of confusion.
Finally, for one who agrees with our opinion on the merits of the SVP advocated
here, we certainly suggest adopting the formulation.\\

\noindent
{\bf From the GSSM perspective, what is R-parity violation ?
Which terms are R-parity violating ? \\
---  the definition of lepton number, and hence R parity, is actually ambiguous;
though there is a clear MSSM limit.} Naively, one can compare the Lagrangian
of the model (mainly the superpotential and the soft SUSY breaking terms) with 
that of the MSSM and call all the extra terms RPV terms.  This is the commonly
adopted terminology. However, while baryon number is still a clearly definite
concept within the GSSM, lepton number is not. This fact may help to appreciate
why may be baryon number  is still be conserved while lepton number is not.

Within the SM, we have lepton
flavor numbers $L_e$, $L_\mu$, and $L_\tau$ unambiguously defined. Lepton
number is then given by $L=L_e+L_\mu+L_\tau$. For instance, the electron
carries one unit of $L_e$, and only particles within the same multiplet carry
the $L_e$ number. The definition carries over to MSSM. In fact,  the MSSM
might better be interpreted as a supersymmetric version of the SM with the 
global symmetries of lepton number(s) and baryon number assumed as a 
fundamental part of the latter rather than ``accidental" consequence of the gauge
symmetries. The GSSM is more like a natural supersymmetric version of the 
SM of gauge interactions. We have seen that in the GSSM, 
the electron, for example, is not contained totally inside any $\hat{L_i}$ superfield 
multiplet. In fact, it is not contained totally inside the $\hat{L_\za}$ superfield 
multiplets. The naive interpretation of R-parity violation mentioned amounts to 
assigning (opposite) $L_e$ numbers only to $\hat{L_1}$ and 
$\hat{E}_1^{\scriptscriptstyle C}$, 
and $L_\mu$ and $L_\tau$ numbers only to $\hat{L_2}$ and 
$\hat{E}_2^{\scriptscriptstyle C}$ and $\hat{L_3}$ and 
$\hat{E}_3^{\scriptscriptstyle C}$, respectively. So, the $\hat{L_i}$'s carry
lepton number but not the $\hat{L_0}$.  This is only under the SVP. Under a
different parametrization, the $\hat{L_i}$ and $\hat{E}_i^{\scriptscriptstyle C}$
pairs each represent even less directly the exact superfields containing the  
corresponding charged leptons, as off-diagonal contributions would be 
everywhere over the fermions mass matrices (in the superfield basis). Hence,
it is even more unappropriate to assign the lepton numbers in the same way.
The terminology of lepton number, or R-parity, violation is mainly used for 
comparing against MSSM features. For the theory model in itself, under
whatever parametrization, such a definition of lepton number(s) certainly 
sounds too arbitrary.  

Of course from the experimental point of view, lepton number or even lepton
flavor numbers can still be unambiguously defined. This is just like the use 
of strangeness. Lepton flavor numbers cease to be theoretically exact
concepts in a model with leptonic or neutrino flavor mixings. Likewise,
lepton number ceases to be theoretically exact concept in GSSM. Just like
we do not talk about baryon (quark) flavor numbers in the theoretical 
discussion of the SM. When one insists on assigning such lepton number(s)
to the GSSM theoretical ingredients (the superfield multiplets), the exact
meaning of the assigned lepton number(s) then differs among different 
parametrizations and differs from the experimental notion. \\
 
\noindent
{\bf What exactly is ``MSSM + RPV trilinear superpotential parameters" ?}\\
--- In the literature, a model of R-parity violation that received a lot of attention
is described as `MSSM + RPV trilinear superpotential parameters". That is to say,
authors assumed the model Lagrangian is given by that of the MSSM amended 
by the addition  of the $\lambda$-, $\lambda^{\!\prime}$-, and 
$\lambda^{\!\prime\prime}$- type couplings terms to
the superpotential. In some cases, only one or two of the three type of couplings
are assumed. In the worst case, that is claimed as the most general Lagrangian 
or superpotential obtainable from the supersymmetrized SM particle spectrum.
Our discussions above should have illustrated clearly the fallacy of such a claim,
or similar ones. The ``model" mostly received attention, apparently, because it is 
simple and very similar to the MSSM. One does not have to worry much about 
changes in the familiar identity of the superfields as there is no new contributions 
to the tree-level mass matrices for the fermions. Restricting from the GSSM 
to such a model really means assuming {\it all} the other ``RPV" terms vanish. 
Under the SVP, the vanishing of the $\mu_i$'s would be enough to keep the
meaning of the Lagrangian terms for such a model the way it was desired. The
other ``RPV" terms may then be neglected in some phenomenological studies
without enforcing the vanishing assumption strictly. Under a parametrization with 
nonzero $\hat{L}_i$ VEVs, the VEVs must also be assumed to vanish, which means 
assumption on the soft SUSY breaking parameters involved in the scalar potential
as well.\\

\noindent
{\bf What exactly is ``MSSM + RPV bilinear (superpotential) parameters" ?}\\
--- Another relative popular version of RPV model is given by admitting only the
bilinear couplings. A theoretical better motivated and clearly defined option
is to obtain such a model from integrating out the heavier superfield(s) that
give rise to a spontaneous breaking of an otherwise present lepton number
symmetry. Without a background lepton number symmetry to begin with, the 
adding of the bilinear terms to MSSM is less of a clear conceptual issue. 
Looking at it from the perspective of our formulation here, one may of course
take the assumption that the trilinear couplings, namely the $\lambda$-, 
$\lambda^{\!\prime}$-, and $\lambda^{\!\prime\prime}$- type couplings
as well as their soft SUSY breaking counterparts all vanish. The further
vanishing of the $B_i$ parameters could be a further assumption. However,
if the parametrization is fixed as the SVP, the clear division between the class
of ``RPV" parameters of the $\lambda$-, $\lambda^{\!\prime}$-, 
and $\lambda^{\!\prime\prime}$- type couplings, and their ``R-parity"
conserving counterparts is lost. The latter are called the MSSM superpotential
terms, and are supposed to be terms giving the SM Yukawa couplings. As we
have discussed above, the $\lambda$- and $\lambda^{\!\prime}$- type couplings
have no contribution to the diagonal SM (or physical) Yukawa couplings only
when they are parameters defined under the SVP flavor bases. Hence, special
care may be needed to handle such a RPV model.


\end{document}